\documentclass[a4paper,review,11pt,nonatbib]{elsarticle}
\journal{arXiv.org}

\usepackage[a4paper,text={16cm,25cm}]{geometry}
\usepackage[labelsep=period]{caption}

\usepackage{array}
\usepackage{booktabs}
\usepackage{amsthm}
\usepackage{amsmath}
\usepackage{amstext}
\usepackage{amsfonts}
\usepackage{graphicx}
\usepackage{bbm}
\usepackage{todonotes}

\makeatletter
\let\c@author\relax
\makeatother
\usepackage[style=authoryear-comp, backend=biber, uniquename=false, maxnames=9, maxcitenames=2, maxbibnames=99, natbib=true, bibencoding=utf8, dashed=false,terseinits=true, giveninits=true, uniquelist=false, doi=false, isbn=false, url=true, sortcites=false, date=year,safeinputenc]{biblatex}

\DeclareFieldFormat{url}{\texttt{\url{#1}}}
\DeclareFieldFormat[article]{pages}{#1}
\DeclareFieldFormat[inproceedings]{pages}{\lowercase{pp.}#1}
\DeclareFieldFormat[incollection]{pages}{\lowercase{pp.}#1}
\DeclareFieldFormat[article]{volume}{\mkbibbold{#1}}
\DeclareFieldFormat[article]{number}{\mkbibparens{#1}}
\DeclareFieldFormat[article]{title}{\MakeCapital{#1}}
\DeclareFieldFormat[inproceedings]{title}{#1}
\DeclareFieldFormat{shorthandwidth}{#1}
\usepackage{xpatch}
\xpatchbibmacro{volume+number+eid}{\setunit*{\adddot}}{}{}{}
\renewbibmacro{in:}{%
  \ifentrytype{article}{}{%
  \printtext{\bibstring{in}\intitlepunct}}}
\makeatletter
\DeclareDelimFormat[cbx@textcite]{nameyeardelim}{\addspace}
\makeatother

\usepackage{algorithm}
\usepackage{algorithmicx}
\usepackage{algpseudocode}

\usepackage{xurl}
\usepackage{hyperref}
\hypersetup{
  colorlinks=true,
  linkcolor=blue,
  citecolor=blue,
  urlcolor=blue}

\usepackage{bm}

\newcommand{\pkg}[1]{{\normalfont\fontseries{b}\selectfont #1}}
\let\proglang=\textsf
\let\code=\texttt

\newcommand{\ie}{i.e.}
\newcommand{\eg}{e.g.}

\usepackage{enumitem}

\begin{document}

\begin{frontmatter}

\title{Distributed ARIMA models for ultra-long time series}

\author[mainaddress]{Xiaoqian Wang}
\ead{xiaoqianwang@buaa.edu.cn}

\author[mainaddress]{Yanfei Kang}
\ead{yanfeikang@buaa.edu.cn}

\author[secondaryaddress]{Rob J Hyndman}
\ead{rob.hyndman@monash.edu}

\author[thirdaddress]{Feng Li\corref{correspondingauthor}}
\cortext[correspondingauthor]{Corresponding author}
\ead{feng.li@cufe.edu.cn}

\address[mainaddress]{School of Economics and Management, Beihang University, Beijing,
  100191, China} \address[secondaryaddress]{Department of Econometrics and Business
  Statistics, Monash University, Clayton, VIC 3800, Australia}
\address[thirdaddress]{School of Statistics and Mathematics, Central University of Finance
  and Economics, Beijing 102206, China}

\begin{abstract}
Providing forecasts for ultra-long time series plays a vital role in various activities, such as investment decisions, industrial production arrangements, and farm management. This paper develops a novel distributed forecasting framework to tackle challenges associated with forecasting ultra-long time series by using the industry-standard MapReduce framework. The proposed model combination approach retains the local time dependency and utilizes a straightforward implementation of splitting across samples to facilitate distributed forecasting by combining the local estimators of time series models delivered from worker nodes and minimizing a global loss function. In this way, instead of unrealistically assuming the data generating process (DGP) of an ultra-long time series stays invariant, we make assumptions only on the DGP of subseries spanning shorter time periods. We investigate the performance of the proposed approach with AutoRegressive Integrated Moving Average (ARIMA) models using the real data application as well as numerical simulations. Compared to directly fitting the whole data with ARIMA models, our approach results in improved forecasting accuracy and computational efficiency both in point forecasts and prediction intervals, especially for longer forecast horizons. Moreover, we explore some potential factors that may affect the forecasting performance of our approach.
\end{abstract}


\end{frontmatter}

\newpage
\section{Introduction}
\label{sec:intro}

Ultra-long time series (\ie~time series data observed over a long time interval) are
becoming increasingly common. Examples include hourly electricity demands spanning several
years, stock indices observed every minute over several months, daily maximum temperatures
recorded for hundreds of years, and streaming data continuously generated in
real-time. Attempts to forecast these data play a vital role in investment decisions,
industrial production arrangement, farm management, and business risk
identification. However, it is challenging to deal with such long time series with
traditional time series forecasting approaches.

We identify three significant challenges associated with forecasting ultra-long time
series. First, the optimization of parameters in training forecasting algorithms is
time-consuming due to the time dependency nature of time series. Second, processing time
series spanning such a long time interval drives significant storage requirements,
especially in the algorithms' training process, where a standalone computer could hardly
tackle. The third and most serious difficulty is that the standard time series models do
not perform well for ultra-long time series \citep{hyndman2018forecasting}. One possible
reason is that it is usually unrealistic to assume that the data generating process (DGP)
of time series has remained invariant over an ultra-long time. Hence, there is an apparent
difference between the models we use and the actual DGP. The more realistic idea is to
assume that the DGP stays locally stable for short time-windows.

Forecasters have made some attempts to address these limitations in forecasting ultra-long
time series. A straightforward approach is to discard the earliest observations and use
the resulting shorter time series for model fitting. But this approach only works well for
forecasting a few future values, and is not an efficient use of the available historical
data. A better approach is to allow the model itself to evolve over time. For example, ARIMA
(AutoRegressive Integrated Moving Average) models and ETS (ExponenTial Smoothing) models can
address this issue by allowing the trend and seasonal components to vary over time
\citep{hyndman2018forecasting}. An alternative, proposed by \citet{das2020predictive}, is
to apply a model-free prediction assuming that the series changes slowly and smoothly with
time. However, the aforementioned methods require considerable computational time in
model fitting and parameter optimization, making them less practically useful in modern
enterprises.

In industry, distributed computing platforms usually lack forecasting
modules. For example, it is well known that Spark supports time series forecasting poorly,
especially multi-step forecasting. To support large-scale time series
forecasting on such platforms, practitioners commonly have to adopt inadequate but available
methods on distributed platforms \citep{meng2016mllib,galicia2018novel}. For instance, one
has to utilize the regression models in Spark's \pkg{MLlib} to implement an autoregressive
type regression and artificially convert the multi-step prediction problem into
multi-subproblems, to fit the Spark framework for time series forecasting with improved
computational efficiency \citep{galicia2018novel}.

On the other hand, the unprecedented scale of data collection has driven a vast literature
on studying statistical estimation and inference problems in a distributed manner, both in
a frequentist setting
\citep{kleiner2014scalable,zhang2015divide,Jordan2019,chen2019quantile}, and a Bayesian
setting
\citep{suchard2010understanding,wang2013parallelizing,maclaurin2015firefly,coluccia2016bayesian}.
Among the existing methods, the divide-and-conquer approach, one of the most important and
easy to implement algorithms on distributed platforms, provides an algorithm design
paradigm in which a given problem is divided into a set of related subproblems that are
simple to process. Their solutions are then aggregated using proper aggregation strategies. Ideally, the subproblems can be solved in parallel, which is computationally more
manageable than dealing with the original problem. Due to its simplicity for parallel
processing, the divide-and-conquer strategy has been widely used in the statistical literature
to untangle large-scale problems with independent data \citep[see, \eg,][and
Section~\ref{sec:tsforecast} for detailed
descriptions]{zhang2013communication,kleiner2014scalable,lee2015communication,zhu2021least,pan2021note}.

In this paper, we follow the idea of divide-and-conquer and provide a novel approach to
addressing the large-scale time series forecasting problems in distributed environments.
Specifically, we propose a distributed time series forecasting framework, in which the
long time series is first divided into several subseries spanning short time periods, and
models can be applied to each subseries under the reasonable assumption that the DGP
remains invariant over a short time. In this view, our framework has the flavor of a
``varying coefficient model'' \citep{fan2008statistical} for a long time series. However,
unlike the varying coefficient models, we combine the local estimators trained on each
subseries using weighted least squares to minimize a global loss function.

Considering a general loss function enables our framework to, in principle, be applied to
statistical time series forecasting models that can be considered as parametric regression
problems. However, properly combining the local estimators is of great importance and challenge in practice.
For example, directly combining the original parameters of ARIMA models trained on subseries
sometimes leads to a global ARIMA model with roots inside the unit circle, thus causing
the stationary, causality, or invertibility problems. Therefore, when applying the proposed
distributed forecasting framework to a specific type of model, the key issue is how to process
local estimators based on the model settings so that they can be combined properly.
Conventionally, ARIMA models are among the most widely used forecasting models because
(i) they can handle non-stationary and seasonal patterns, (ii) ARIMA models also frequently
serve as benchmark methods because of their excellent performance
\citep{montero2020fforma,wang2021uncertainty}.
Nonetheless, such models are difficult to scale up with the current Spark distributed platform
due to the nature of time dependency, making it infeasible for large scale time series
forecasting. Therefore, we restrict our attention to ARIMA models and propose
the \emph{Distributed} ARIMA (DARIMA) models, in which a necessary linear transformation step
is involved to address the aforementioned issue of estimator combination.

The proposed method preserves the local time dependency and performs a straightforward
achievement of splitting across samples to make distributed forecasting possible for
ultra-long time series with only one round of communication. Thus, the proposed method
can be naturally integrated into industry-standard distributed systems with a
MapReduce architecture. Such a MapReduce algorithm requires only one ``master-and-worker''
communication for each worker node and avoids further iterative steps. No direct communication
between workers is required. To this end, it is highly efficient in terms of communication.
Additionally, our proposed method is scalable for a large forecast horizon, which is the
typical case when dealing with ultra-long time series, and is therefore superior to the
distributed forecasting methods (reviewed in Section~\ref{sec:tsforecast}) in which multiple
trainings or iterations are required in order to perform multi-step forecasting.

Our \emph{Distributed} ARIMA modeling framework is built using an efficient
distributed computing algorithm without modifying the underlying estimation scheme for
individual time series models, making it possible to incorporate a large variation of
forecasting models. In both the real data application and the simulations, we show that
our approach consistently achieves an improved forecasting accuracy over conventional global
time series modeling approaches, both in point forecasts and prediction intervals. The
achieved performance improvements become more pronounced with increasing forecast
horizon. Moreover, our approach delivers substantially improved computational efficiency
for ultra-long time series.

The rest of the paper is organized as follows. Section~\ref{sec:review} describes the
distributed systems, ARIMA models, and highlights of our contributions to ultra-long time
series forecasting. Section~\ref{sec:method} introduces the framework of the proposed
forecasting approach in distributed systems, further elaborated by its core
components. Section~\ref{sec:application} applies the proposed method with a real data and
provides a sensitivity analysis. A simulation study is performed in Section~\ref{sec:sim}
to further investigate and justify our proposed method in terms of forecasting accuracy
and computational cost. Section~\ref{sec:discussion} discusses other potentials
and suggests possible avenues of research. Finally, Section~\ref{sec:conclusion} concludes
the paper.

\section{Background}
\label{sec:review}

This section surveys the forecasting challenges on distributed systems with a special
focus on the ARIMA models, and highlights the contributions of our framework.

\subsection{Forecasting with Distributed Systems}
\label{sec:tsforecast}

A typical distributed system consists of two core components: the Distributed File System
(DFS) and the MapReduce framework. See \ref{sec:distr} for an overview for distributed
systems. The DFS provides the primary storage infrastructure for distributed systems. By
storing files in multiple devices, the DFS effectively eliminates the negative impacts of
data loss and data corruption. It enables devices to handle large-scale data sets and
access data files in parallel. MapReduce provides the batch-based computing framework for
distributed systems. The MapReduce framework refers to two steps, namely Map and
Reduce. The input data set is first split into a collection of independent data tuples,
represented by key/value pairs. The Map task takes the assigned tuples, processes them in
a parallel manner, and creates a set of key/value pairs as the output, illustrated by
$\langle k1, v1 \rangle \rightarrow Map(\cdot) \rightarrow list(\langle k2,
v2\rangle)$. Subsequently, the Reduce task takes the intermediate outputs that come from
the Map tasks as its inputs, combines these data, and produces a new set of key/value
pairs as the output, which can be described as
$\langle k2, list(v2) \rangle \rightarrow Reduce(\cdot) \rightarrow list(\langle k3, v3
\rangle)$. The main advantage of the MapReduce computing paradigm is that data processing
is enabled to be easily extended on multiple computing nodes with high computational
efficiency.

Many distributed systems were designed for processing massive independent data. However,
a distinct feature of time series data is that the observations are serially dependent,
and so additional considerations are required in processing time series with distributed
systems. How to efficiently bridge time series forecasting with the distributed systems
is of crucial importance in the forecasting domain.

We identify four main challenges associated with using distributed systems for time
series forecasting: (i) time series partitioning for MapReduce tasks
\citep{li2014rolling}; (ii) independent subseries modeling with distributed systems;
(iii) model communication between worker nodes; and (iv) distributed time series
forecasting, especially for multi-step prediction \citep{galicia2018novel}.

Some attempts have been made by researchers to allow for efficient implementations of
large-scale time series analysis. For example, \citet{mirko2013hadoop} propose a
Hadoop-based analysis framework for efficient data collection and accurate preprocessing,
which are at least as important as the algorithm selection for a given forecasting problem.
Unlike \citet{mirko2013hadoop}, \citet{li2014rolling} focus on model selection.
Specifically, MapReduce is used to perform cross-validation and facilitate parallel
rolling-window forecasting using the training set of a large-scale time series data. The
forecasting errors are then collected and used to calculate the accuracy, allowing model
comparison and selection. Therefore, the method is not designed to address the challenges
associated with forecasting using ultra-long historical data. \citet{talavera2018big} and
\citet{galicia2018novel} manage multi-step forecasting for ultra-long time series from the
perspective of machine learning by computing with the Spark platform. Specifically,
\citet{talavera2018big} require $H$ iterations to perform multi-step forecasting, while
\citet{galicia2018novel} formulate the forecasting problem as $H$ parallel forecasting
subproblems to support multivariate regression with the \pkg{MLlib} library, where $H$ is the
forecast horizon. Consequently, the approaches they propose have poor scalability for a large
forecast horizon. However, ultra-long time series forecasting is typically characterized with the
need to forecast quite a few future values in practice. A recent overview of forecasting with
big data time series is provided in Section 2.7 of the encyclopedic review in
\citet{petropoulos2020forecasting}. In summary, the attempts mentioned above provide a premise
for distributed time series forecasting but it is still impractical to apply these methods to
handle ultra-long time series for the direct purpose of forecasting quite a few future
observations.

In recent years, research has been deeply engaged in distributed statistical inference
with concerns about practical computational cost for large-scale data processing. The most
prominent papers build upon the divide-and-conquer strategy.
The communication cost, which refers to the time cost for data communication between different computer
nodes, is often recognized as one of the key challenges faced by distributed statistical inference.
The first stream of research
employs one-shot aggregation and averages the local estimators computed on each batch of
the entire data to obtain the global estimators. Examples include but are not limited to
the subsampled average algorithm based on an additional level of sampling on worker nodes
\citep{zhang2013communication}, the generative model parameters estimation via the maximum
likelihood estimator \citep{liu2014distributed}, quantile regression processes
\citep{volgushev2019distributed}, parametric regression handled by least square
approximation \citep{zhu2021least}, and nonparametric and infinite-dimensional regression
\citep{zhang2015divide}.

Another stream of research looks at the underlying iterative algorithms with multi-round
communications and aggregations for distributed optimization. Multiple iterations are
considered with the purpose of further matching the aggregated estimators to the global
estimators. These iterations unfortunately result in considerable ``master-and-worker''
communication, which is communicationally expensive.
For example, \citet{shamir2014communication} propose a distributed
approximate Newton algorithm to solve general subproblems available locally before
averaging computations, requiring a constant number of iterations (and thus rounds of
communication). Based on their framework, recent works by \citet{wang2017efficient} and
\citet{Jordan2019} propose iterative methods with communication efficiency for distributed
sparse learning and Bayesian inference. \citet{chen2019quantile} restrict themselves to
refine the estimator of quantile regression via multiple rounds of aggregations under
distributed computing environment. In addition, another popular approach is the
alternating direction method of multipliers \citep[ADMM,][]{boyd2011distributed} developed
for distributed convex optimization. It blends the decomposability of dual ascent with the
superior convergence properties of the method of multipliers.

Both the streams are difficult to apply directly to time series forecasting models. The
one-shot averaging strategy is straightforward to implement and requires only a single
round of communication. While naively merging the local estimators that are processed
separately may yield inference procedures that are highly biased and variable, leading to
inefficient estimations in most occasions; see \citet{zhang2013communication},
\citet{shamir2014communication}, \citet{Jordan2019}, and \citet{pan2021note} for further
discussions. The divide-and-conquer strategy is difficult to the time series
forecasting. So some research focuses on distributed learning by splitting across the
features in the frequency domain for different time series rather than samples (different
timestamps) \citep{sommer2021online,gonccalves2021critical}. Distributed algorithms like
ADMM suffer a crucial limitation that (i)~they require a reimplementation of each
estimation scheme with distributed systems, and (ii)~they can be very slow to converge to
high accuracy compared to existing algorithm designed for standalone computers, see
\citet{boyd2011distributed} for more details. Additionally, communication cost is
recognized as a key challenge faced by statistical computation on a distributed system
\citep{Jordan2019,zhu2021least}.

The studies reviewed above focus on scaling up optimization algorithms for solving
large-scale statistical tasks in a distributed manner, while our study adopts a model
combination strategy in which model parameters obtained from multiple subseries are
aggregated to minimize the global loss function. In this way, our proposed method can achieve
lower computational and implementation complexity than modeling the whole ultra-long time
series with a single model.

\subsection{ARIMA Models}
\label{sec:arima}

An ARIMA (AutoRegressive Integrated Moving Average) model is composed of differencing,
autoregressive (AR), and moving average (MA) components \citep{box2015time}. We refer to
an ARIMA model as ARIMA$(p, d, q)$, where $p$ is the order of the AR component, $d$ is the
number of differences required for a stationary series, and $q$ is the order of the MA
component. An ARIMA model can be extended to a seasonal ARIMA model by including
additional seasonal terms to deal with time series exhibiting strong seasonal behavior. A
seasonal ARIMA model is generally denoted as ARIMA$(p, d, q)(P, D, Q)_m$, where the
uppercase $P, D, Q$ refer to the AR order, the number of differences required for a
seasonally stationary series, and the MA order for the seasonal component, respectively,
while $m$ denotes the period of the seasonality \citep{tsay2005analysis}.

An ARIMA$(p, d, q)(P, D, Q)_m$ model for time series $\{y_{t}, t \in \mathbb{Z}\}$ can be
written with the backshift notation as
\begin{align}
  \label{eq:sarima_raw}
  \bigg(1-\sum_{i=1}^{p}\phi_{i} B^{i}\bigg) \bigg(1-\sum_{i=1}^{P}\Phi_{i}
  B^{im}\bigg)(1-B)^{d}(1-B^{m})^{D} y_{t}\\ \nonumber
  =\bigg(1+\sum_{i=1}^{q}\theta_{i} B^{i}\bigg)\bigg(1+\sum_{i=1}^{Q}\Theta_{i} B^{im}\bigg) \varepsilon_{t},
\end{align}
where $B$ is the backward shift operator, $\varepsilon_{t}$ is white noise, $m$ is the
length of the seasonal period, $\phi_{i}$ and $\Phi_{i}$ refer to the AR parameters of
non-seasonal and seasonal parts, $\theta_{i}$ and $\Theta_{i}$ refer to the MA parameters
of non-seasonal and seasonal parts respectively.

Combinations of the non-seasonal orders $p, d, q$ and seasonal orders $P, D, Q$ provide a
rich variation of ARIMA models. Consequently, identifying the best model among these
possibilities is of crucial importance in obtaining good forecasting performance using
ARIMA models. Fortunately, vast automatic ARIMA model selection schemes are developed. One
of the most widely used algorithms is the \code{auto.arima()} function developed for
automatic time series forecasting with ARIMA models in the \proglang{R} package
\pkg{forecast} \citep{Hyndman2008b}. Despite that those algorithms allow us to implement
the order selection process with relative ease in a standalone computer for short time
series, efficient ARIMA model selection for ultra-long time series is challenging with
modern distributed computational environments.

\begin{figure}
    \centering
    \includegraphics[width=\textwidth]{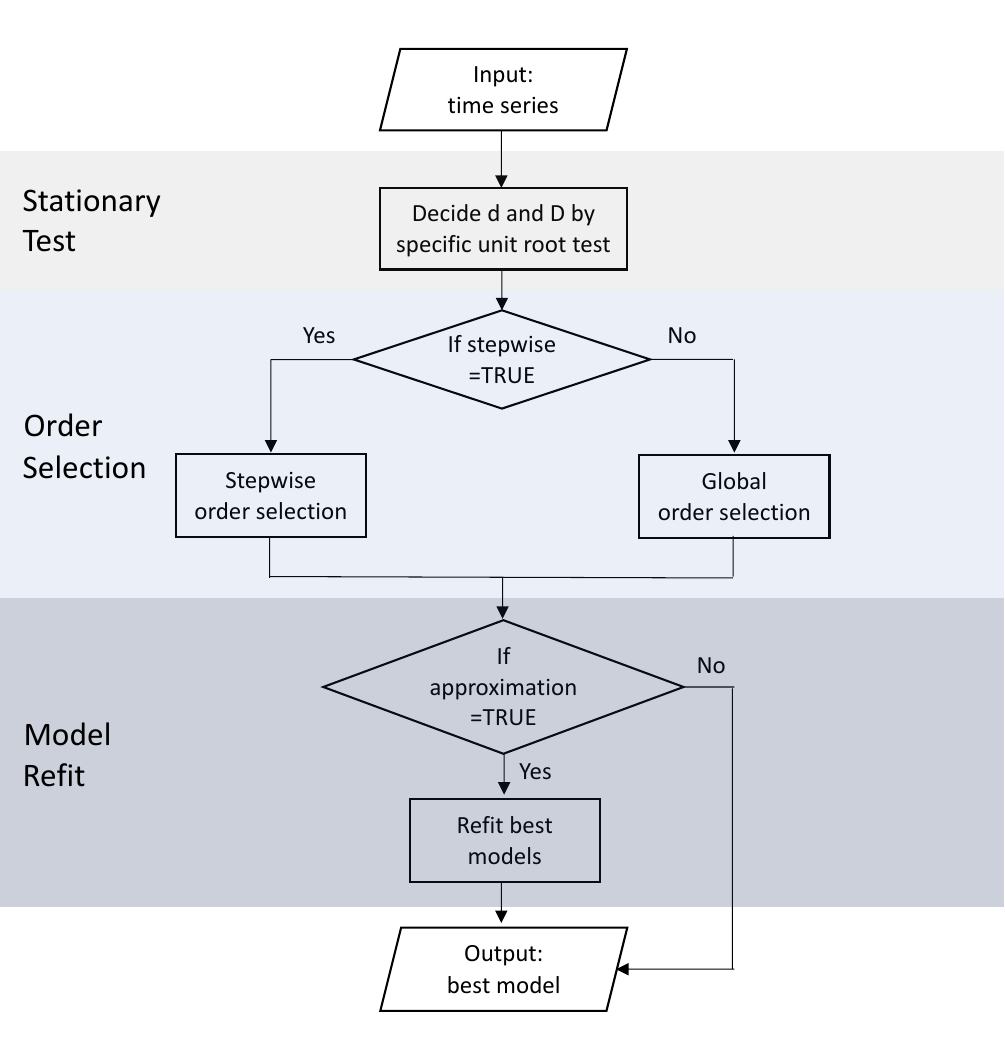}
    \caption{The procedure of an automatic ARIMA forecasting framework, taking the
      \code{auto.arima()} algorithm as an example.}
    \label{fig:auto.arima}
\end{figure}

We take the algorithm in the \code{auto.arima()} function to describe the ARIMA model
selection process. Other algorithms follow a similar fashion. Figure~\ref{fig:auto.arima}
depicts how the \code{auto.arima()} function is applied to conduct a search process over
possible models. The algorithm consists of three main steps: stationary tests, order
selection, and model refit. First, the stationary tests aim to decide the order of
first-differencing and seasonal-differencing, using a KPSS test
\citep{kwiatkowski1992testing} for estimating $d$ and either a Canova-Hansen test
\citep{canova1995seasonal} or a measure of seasonality \citep{hyndman2018forecasting} for
estimating $D$. Second, the order selection process chooses the model orders via an
information criterion such as AIC, AICc, or BIC values. There are two options for the
order selection approach, namely (greedy) stepwise selection and global selection, which
can be customized according to time series characteristics such as time series length and
seasonality. Such selection can be time-consuming because each information criterion is
obtained by a model fitting process. Finally, the selected model orders are applied to
refit best models without approximation if the information criteria used for model
selection are approximated.

The automatic ARIMA modeling has been extended in many ways by forecasting researchers
\citep[\eg ,][]{calheiros2014workload, shang2017grouped, makridakis2020m4}. Despite its
superb performance in forecasting time series, several difficulties hinder the extension
of this approach to ultra-long time series. We use the electricity demand for the
Northeast Massachusetts (NEMASSBOST) zone of the Global Energy Forecasting Competition
2017 \citep[GEFCom2017;][]{hong2019global} to elaborate on the following challenges of
extending the \code{auto.arima()} function to ultra-long time series data.

\begin{figure}
    \centering
    \includegraphics[width=\textwidth]{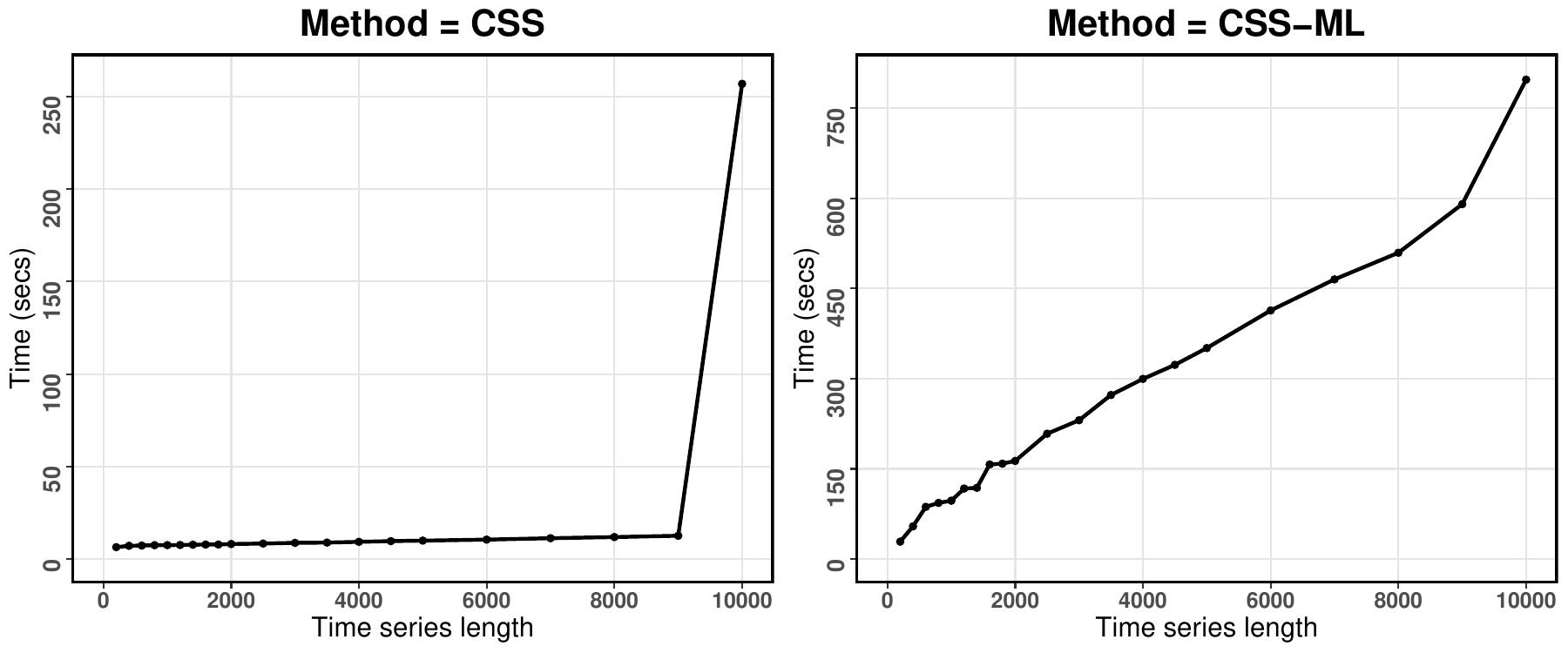}
    \caption{The effect of time series length and fitting method on the execution time of
      time series modeling with the ARIMA model as an example, with other arguments as the
      default setting. The electricity demand series for NEMASSBOST zone of GEFCom2017 is
      used.}
    \label{fig:lentime}
\end{figure}

\begin{enumerate}
\item Modeling the whole time series with a single model relies on an unrealistic
  assumption that the DGP of time series has remained invariant over an ultra-long
  period. Note that the assumption is made on the DGP of the target time series, rather
  than the time series itself.

\item Order selection is an extremely time-consuming process, which requires to fit all
  available models. Even though we can select model orders by adopting global order
  selection approach with parallelism, it still takes a lot of time to run a single time
  series model for ultra-long time series. The computational time grows exponentially with
  the length of time series increasing.

\item Multiple models may be considered in the model refit process because the
  \code{auto.arima()} function carries out several strict checks for unit roots, also
  resulting in a loss of computing efficiency.

\item The existing approaches for model fitting, such as CSS (conditional sum-of-squares),
  ML (maximum likelihood), and hybrid CSS-ML (find starting values by CSS, then ML), are
  hard to parallel due to the nature of time dependency. The ML approach is the most
  commonly used but time-consuming approach for fitting ARIMA models
  \citep{Hyndman2008b}. Figure~\ref{fig:lentime} compares the execution time of the
  \code{auto.arima()} function under the CSS and CSS-ML fitting methods, and shows the
  impact of fitting methods on the function's execution time.
  When considering minimizing the conditional sum-of-squares as the estimation method
  for ARIMA models, the \code{auto.arima()} function in the \pkg{forecast} package for
  \proglang{R} commonly uses the Broyden–Fletcher–Goldfarb–Shanno (BFGS) algorithm
  for optimization purposes. The computational complexity of ARIMA modeling is
  $\mathcal{O}(n^{2}T)$ when the non-seasonal orders ($p, d, q$) and seasonal orders
  ($P, D, Q$) are given, where $n$ is the number of parameters (\ie~$n=p+q+P+Q$),
  and $T$ is the length of the time series of interest.

\item The length of time series has a significant impact on automatic ARIMA modeling. We
  notice that a standalone computer may not have sufficient resources to fit an ultra-long
  time series. From Figure~\ref{fig:lentime}, we find that time series with longer length
  yield much longer computation time, which provides another good explanation of why the
  order selection and model refit processes are time-consuming.

\item Most model selection schemes only allow a small range of lag values in ARIMA
  modeling to limit the computation. The maximum values of model orders directly determine
  the available models in the order selection process. If the model orders are allowed to
  take a broader range of values, the number of candidate models will increase
  rapidly. Therefore, relaxing the restriction of maximum values of model orders becomes
  an obstacle in automating ARIMA modeling for ultra-long time series.

\end{enumerate}

In this study, the proposed algorithm is designed to tackle these challenges by estimating ARIMA
models on distributed systems, making it suitable for the processing of big data time series.
More specifically, the proposed algorithm retains the local time dependency and utilizes a straightforward
implementation of splitting across samples to facilitate distributed forecasting for
ultra-long time series with only one round of communication. To the best of our knowledge,
this study is the first distributed forecasting approach that integrates distributed
computing and forecast combinations to process time series spanning large-scale time
periods, in which we use weighted least squares to combine the local estimators trained on
each subseries by minimizing a global loss function. The proposed framework helps in
extending existing forecasting methods to distributed ultra-long time series forecasting
by merely making assumptions about the DGP of subseries spanning a short time
interval. Our proposed approach makes the following contributions compared to the existing
literature and implementations.

\begin{enumerate}
\item We extend the distributed least-square approximation \citep[DLSA,][]{zhu2021least}
  method, which is designed to solve pure regression-type problems with observable
  covariates, to address the challenges associated with forecasting ultra-long time series
  in a distributed manner. While the theoretical and empirical work by
  \citet{zhu2021least} guarantees the statistical properties for independent data.

\item Although the DLSA method ensures the estimators being consistent with the global
  model, where all data are used to fit a single model, this is not our only concern with
  time series forecasting because a conventional time series model usually unrealistically
  assumes the DGP of an ultra-long time series is invariant. In our paper, the DGPs of
  each subseries are allowed to vary over time, thus enabling the possible evolution of
  trend and seasonality across consecutive subseries. With the support of DLSA, the local
  estimators computed on each subseries are aggregated using weighted least squares in
  order to minimize the global loss function. This further prevents overfitting as a
  result of averaging multiple models rather than selecting a single model.

\item Compared with algorithm level parallelization, such as \citet{boyd2011distributed},
  our framework is intuitive and easy to implement. In principle, many
  forecasting models are possible to elaborate with our framework. Moreover, our framework
  outperforms competing methods with improved computational efficiency, especially for
  long-term forecasting, which is necessary for many fields, such as investment decisions,
  industrial production arrangements, and farm management.

\item Time series models, such as the ARIMA model or GARCH models, also model the error
  terms in a parametric form. Our study shows that directly applying the DLSA, which
  focuses on the coefficients of regression type problems, to all the parameters in time
  series models may cause the stationary, causality, or invertibility problems. Our work
  further tackles this issue by a necessary ARIMA transformation step.

\item Our approach retains a solid theoretical foundation and our proposed scheme can also
  be viewed as a model combination approach in the sense that it combines model parameters
  from the multiple subseries, in contrast to the classic forecast combinations of
  different forecasting methods \citep[\eg
  ,][]{montero2020fforma,kang2020gratis,LI2020imaging,wang2021uncertainty}.

\end{enumerate}

\section{Distributed Forecasting for Ultra-long Time Series}
\label{sec:method}

Given a time series spanning a long stretch of time, $\{y_{t}; t=1, 2, \dots, T\}$, we aim
to develop a novel framework to work well for forecasting the future $H$
observations. Define $\mathcal{S}=\{1, 2, \dots, T\}$ to be the timestamp sequence of time
series $\{y_{t}\}$. Then the parameter estimation problem can be formulated as
$f( \theta ,\Sigma \mid y_t, t \in \mathcal{S})$, where $f$ is a parameter estimation
algorithm, $\theta$ denotes the global parameters, and $\Sigma$ denotes the covariance
matrix for the global parameters.

\begin{figure}
    \centering
    \includegraphics[width=\textwidth]{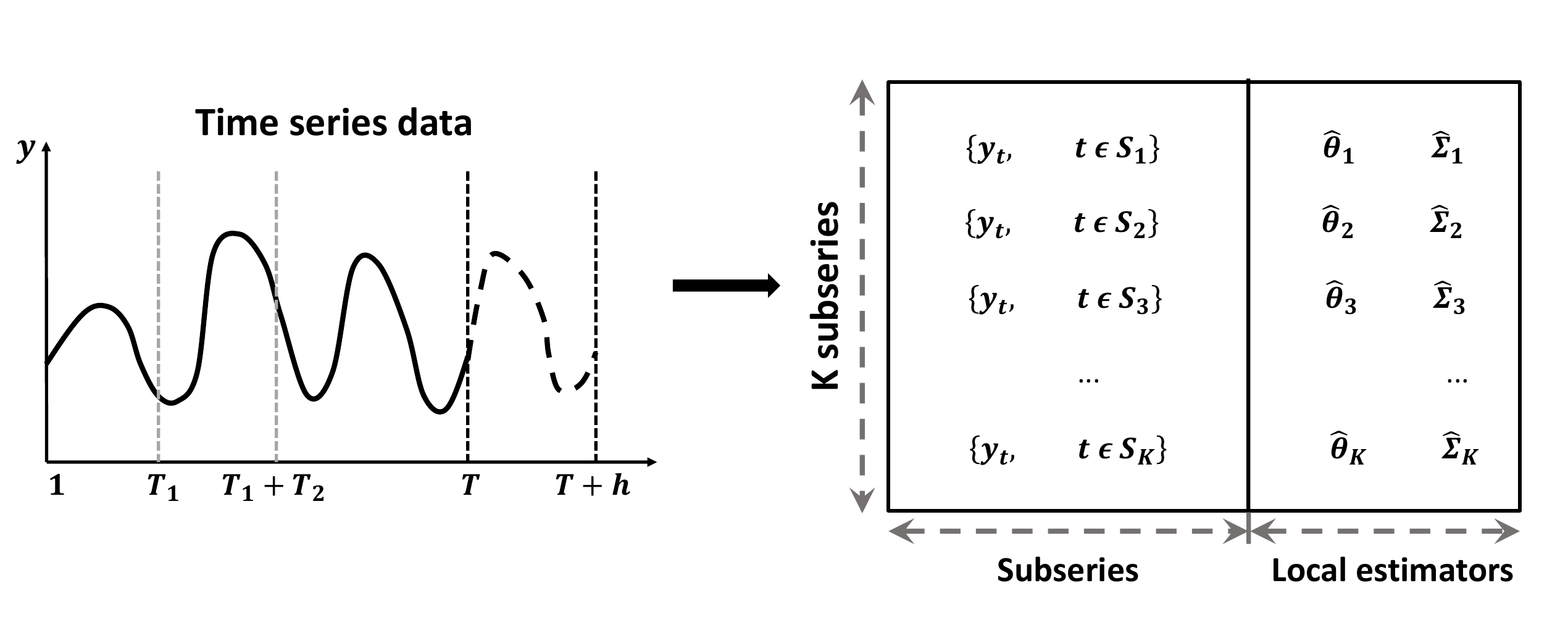}
    \caption{Illustration of forecasting problem and time series split.}
    \label{fig:split}
\end{figure}

Nevertheless, the above statement relies on the assumption that the underlying DGP of the
time series remains the same over a long stretch of time, which is unlikely in
reality. Alternatively, suppose the whole time series is split into $K$ subseries with
contiguous time intervals; that is $\mathcal{S}=\cup_{k=1}^{K} \mathcal{S}_{k}$, where
$\mathcal{S}_{k}$ collects the timestamps of $k$th subseries. We know that
$T = \sum_{k=1}^{K}T_k$, where $T_k$ is the length of the $k$th subseries. Consequently,
we divide an ultra-long time series into several subseries with a realistic assumption
made about the DGP of each subseries, as illustrated in Figure~\ref{fig:split}. In this
way, the parameter estimation problem is transformed into $K$ subproblems and one
combination problem as follows:
\begin{align*}
f(\theta ,\Sigma \mid y_t, t \in \mathcal{S}) = g\big( f_1( \theta_1 ,\Sigma_1 \mid y_t, t \in \mathcal{S}_1), \ldots, f_K( \theta_K ,\Sigma_K \mid y_t, t \in \mathcal{S}_K) \big),
\end{align*}
where $f_k$ denotes the parameter estimation problem for the $k$th subseries, and $g$ is a
combination algorithm applied to combine the local estimators of subseries. Here we are
combining the models before forecasting, rather than forecasting from each model and then
combining the forecasts \citep{kang2020gratis,wang2021uncertainty}. In the simplest
situation, $g(\cdot)$ could be just a single mean function, and our framework could be
viewed as a model averaging approach. Note that the optimal length of split subseries can
be different for different time series. One has to balance the computational resources
with the needed length to replicate the underlying structure in the time series. For
example, the subseries of an hourly time series should be long enough to capture the daily
pattern, while the monthly pattern can be ignored due to the limited computational
resources and processed by preprocessing steps such as time series decomposition before
distributed computing.

\begin{figure}
    \centering
    \includegraphics[width=\textwidth]{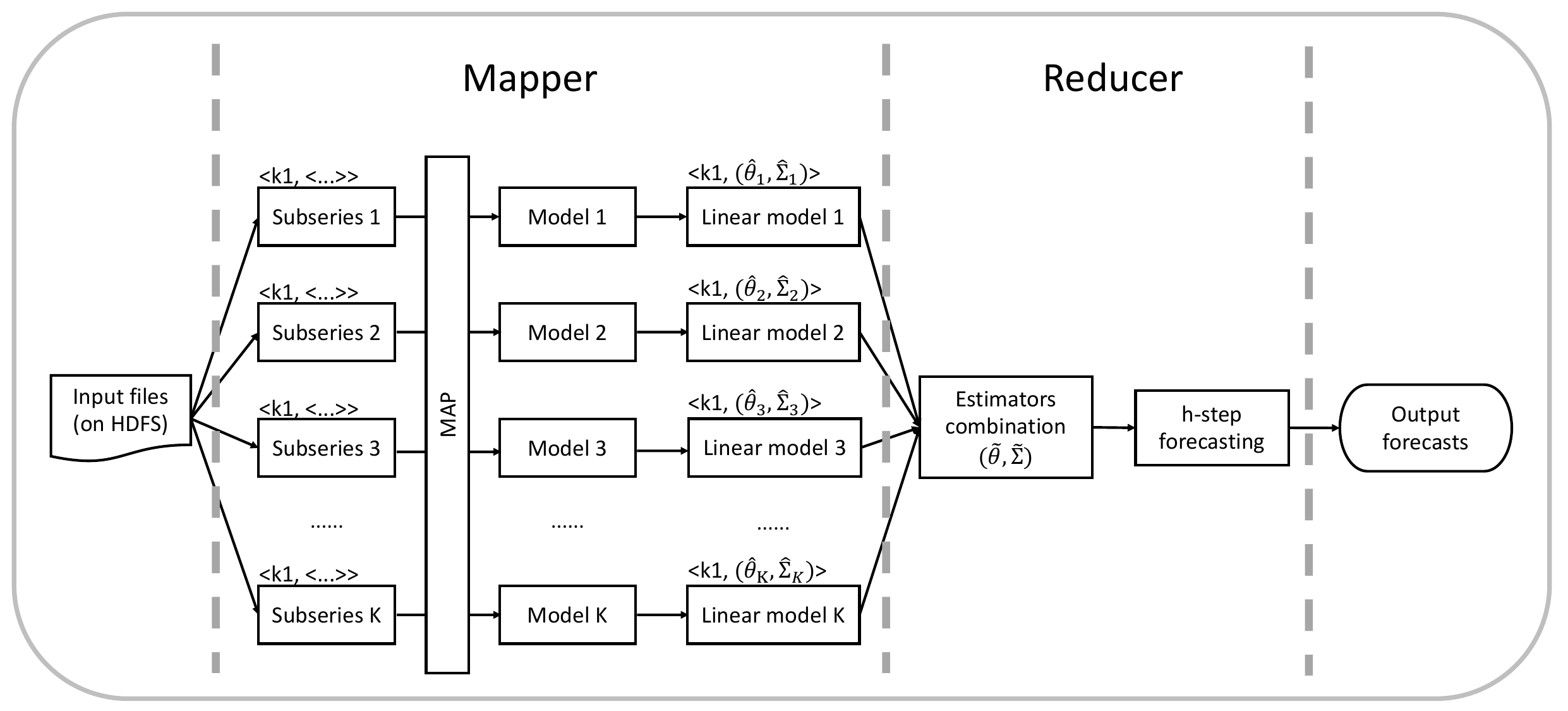}
    \caption{The proposed framework for time series forecasting on distributed systems.}
    \label{fig:framework}
\end{figure}

Figure~\ref{fig:framework} outlines the proposed framework to forecast an ultra-long time
series on distributed systems. We assume that the historical observations and their
timestamps are stored in the DFS before being processed by our framework. We document the
pseudocode for the Mapper and Reducer of the proposed approach in \ref{sec:dfalg}.
Such a MapReduce algorithm requires only one ``master-and-worker'' communication for each worker
node and avoids further iterative steps required by several distributed methods,
see Section~\ref{sec:tsforecast} for further discussions.
Hence, our proposed method is highly efficient in terms of communication. In simple
terms, the proposed framework consists of the following steps.

\begin{enumerate}[noitemsep, leftmargin=*,labelindent=16pt, label=\bfseries Step
  \arabic*:]

\item \textbf{Preprocessing}. Split the whole time series into $K$ subseries, as shown in
  Figure~\ref{fig:split}, which is done automatically with distributed systems.

\item \textbf{Modeling}. Train a model for each subseries via worker nodes by assuming
  that the DGP of subseries remains the same over the short time-windows.

\item \textbf{Linear transformation}. Convert the trained models in Step 2 into $K$ linear
  representations described in Section~\ref{sec:arima2ar}.

\item \textbf{Estimator combination}. Combine the local estimators obtained in Step 3 by
  minimizing the global loss function described in Section~\ref{sec:dlsa}.

\item \textbf{Forecasting}. Forecast the next $H$ observations by using the combined
  estimators described in Sections~\ref{sec:point} and \ref{sec:pis}.

\end{enumerate}

In this paper, we illustrate our approach with the ARIMA model. Since we split the time series
of interest into $K$ subseries with contiguous time intervals, the computational complexity
of modeling ARIMA for each subseries is reduced to $\mathcal{O}(n^{2}T/K)$ when the
model orders are specified. As a result, when forecasting an ultra-long time series with
an extremely large $T$, our proposed method is computationally more efficient than ARIMA,
which has a computational complexity of $\mathcal{O}(n^{2}T)$, because it solves the large-scale
computation problem in a distributed fashion.

The rest of this section elaborates on the steps and approaches of the
framework. Section~\ref{sec:arima2ar} provides the details of how to convert a general
ARIMA model into a linear representation. Section~\ref{sec:dlsa} entails solving the
problem of combining the local estimators of subseries' models, while
Section~\ref{sec:point} and Section~\ref{sec:pis} describe the multi-step point and
interval forecasting respectively.

\subsection{Linear Representations of ARIMA Models}
\label{sec:arima2ar}

The order selection process identifies the model with the minimum specified information
criterion for each split subseries by using the automatic ARIMA modeling implemented in
the \pkg{forecast} package for \proglang{R} \citep{Hyndman2008b}. Moreover, several checks
are carried out to avoid an output ARIMA model with roots inside the unit circle, thus
ensuring time series properties such as stationarity, causality, and invertibility.
Employing distributed systems to forecast ultra-long time series requires the local models
fitted on the subseries capable of being combined to result in the global model for the
whole series. However, directly combining the original parameters of ARIMA models trained
on subseries may sometimes lead to a global ARIMA model with roots inside the unit circle.
As a result, directly combining all parameters to produce an auxiliary global ARIMA model
could be ill-behaved, thus resulting in numerically unstable forecasts. Consequently, in
this subsection, we are devoted to converting a general ARIMA model into a linear
representation which can be regarded as a regression problem to facilitate the
parameters-based combination.

Following Equation~\eqref{eq:sarima_raw}, a general seasonal ARIMA model with intercept,
time trend, and covariates used in the \pkg{forecast} \proglang{R} package is formally
given by
\begin{align}
\label{eq:sarima}
\bigg(1-\sum_{i=1}^{p}\phi_{i} B^{i}\bigg) \bigg(1-\sum_{i=1}^{P}\Phi_{i} B^{im}\bigg)(1-B)^{d}(1-B^{m})^{D} \bigg(y_{t} - \mu_0 - \mu_1 t - \sum_{j=1}^{l}\eta_{j}\gamma_{j,t}\bigg) \nonumber \\
=\bigg(1+\sum_{i=1}^{q}\theta_{i} B^{i}\bigg)\bigg(1+\sum_{i=1}^{Q}\Theta_{i} B^{im}\bigg) \varepsilon_{t},
\end{align}
where $\mu_0$ is the intercept, $\mu_1$ is called the slope of the linear time trend,
$\gamma_{j,t} (j=1,2,\cdots,l)$ is a covariate at time $t$ and $\eta_{j}$ is its
coefficient. The automatic ARIMA modeling provides flexibility in whether to include the
intercept and time trend terms. The time trend coefficient may be non-zero when $d+D=1$,
while the intercept may be non-zero when $d+D=0$.

Let $x_t = y_{t} - \mu_0 - \mu_1 t - \sum_{j=1}^{l}\eta_{j}\gamma_{j,t}$. Then the
seasonal ARIMA model for time series $\{y_{t}, t \in \mathbb{Z}\}$ is transformed into a
seasonal ARIMA model for time series $\{x_{t}, t \in \mathbb{Z}\}$ without the intercept,
time trend, and covariates terms. First, we convert the seasonal ARIMA model into a
non-seasonal ARMA model.
By using the polynomial multiplication, we assume
that the converted non-seasonal ARMA model is denoted ARMA$(u,v)$, where $u=p+mP+d+mD$ and
$v=q+mQ$. The (possibly non-stationary) ARMA$(u,v)$ is defined as
\begin{align}
\label{eq:arma}
\bigg(1-\sum_{i=1}^{u}\phi_{i}^{\prime} B^{i}\bigg) x_{t}=\bigg(1+\sum_{i=1}^{v}\theta_{i}^{\prime} B^{i}\bigg) \varepsilon_{t},
\end{align}
where $\phi_{i}^{\prime}$ and $\theta_{i}^{\prime}$ refer to the AR and MA parameters
respectively. The converted ARMA model after polynomial multiplications still satisfies
the invertibility condition because of the unit root checks performed in the original
ARIMA modeling, so that all roots of the MA characteristic polynomial lie outside the unit
circle.

The next task involves converting the ARMA$(u,v)$ model for time series $\{x_{t}\}$ to its
linear representation.
Given polynomials
$\phi^{\prime}(B) = \left(1-\sum_{i=1}^{u}\phi_{i}^{\prime} B^{i}\right)$, and
$\theta^{\prime}(B) = \left(1+\sum_{i=1}^{v}\theta_{i}^{\prime} B^{i}\right)$ with roots
outside the unit circle, we have
\begin{align*}
\pi(B)x_t = \frac{\phi^{\prime}(B)}{\theta^{\prime}(B)} x_t = \varepsilon_{t},
\end{align*}
where $\pi (B) = \left(1-\sum_{i=1}^{\infty}\pi_{i} B^{i}\right)$. The parameters of the
converted AR($\infty$) model can be obtained by a recursion process. Consequently, the
linear representation of the original seasonal ARIMA model in Equation~\eqref{eq:sarima}
is given by
\begin{equation}
\label{eq:ar}
y_t = \beta_0 + \beta_1 t + \sum_{i=1}^{\infty}\pi_{i}y_{t-i} + \sum_{j=1}^{l}\eta_{j}\bigg(\gamma_{j,t}-\sum_{i=1}^{\infty}\pi_{i}\gamma_{j,t-i}\bigg) + \varepsilon_{t},
\end{equation}
where
$$
\beta_0 = \mu_0 \bigg( 1- \sum_{i=1}^{\infty}\pi_{i} \bigg) + \mu_1 \sum_{i=1}^{\infty}i\pi_{i}
\qquad \text{and}\qquad
\beta_1 = \mu_1 \bigg( 1- \sum_{i=1}^{\infty}\pi_{i} \bigg).
$$
Thus, the infinite order in the AR representation can be approximated by a large order of
$p^{*}$ to make the AR$(p^{*})$ model infinitely close to the true AR process.

\subsection{The Distributed Least Squares Approximation Method}
\label{sec:dlsa}

Suppose we have obtained the appropriate models for individual subseries by traversing the
model space. The next stage entails solving the problem of combining the local estimators
of each subseries model to perform multi-step forecasting.
Inspired by \citet{zhu2021least}, we aim to solve the time series modeling problem with
the dependency structure. The local ARIMA models trained for the subseries are unified
into the AR representations with a large order, making it possible to estimate the global
parameters in the master node by combining the local estimators delivered from a group of
worker nodes.

Let the model parameters in Equation~\eqref{eq:ar} be given by
$\theta = (\beta_0, \beta_1, \pi_1, \pi_2, \cdots, \pi_p, \eta_{1}, \cdots,
\eta_{l})^{\top}$. If $\mathcal{L}(\theta; y_t)$ is a twice differentiable loss function,
we have the global loss function given by
$\mathcal{L}(\theta)=T^{-1} \sum_{t=1}^{T} \mathcal{L}(\theta ; y_{t})$ and the local loss
function for the $k$th subseries given by
$\mathcal{L}_{k}(\theta)=T_{k}^{-1} \sum_{t \in \mathcal{S}_{k}} \mathcal{L}(\theta;
y_{t})$. By using Taylor's theorem and the relationship between the Hessian and covariance
matrix for Gaussian random variables \citep{yuen2010bayesian}, we have
\begin{align}
\label{eq:loss}
\mathcal{L}(\theta) &=\frac{1}{T} \sum_{k=1}^{K} \sum_{t \in \mathcal{S}_{k}} \mathcal{L}(\theta ; y_{t})
\approx \sum_{k=1}^{K} (\theta-\widehat{\theta}_{k})^{\top} \left(\frac{T_k}{T} \widehat{\Sigma}_{k}^{-1}\right) (\theta-\widehat{\theta}_{k})+c_{2},
\end{align}
where $\widehat{\theta}_{k}$ is the minimizer of the local loss function. That is $\widehat{\theta}_{k} = \arg \min \mathcal{L}_{k}(\theta)$, $c_{1}$ and $c_{2}$ are constants, and $\widehat{\Sigma}_{k}$ is the covariance estimate for local estimator of the $k$th subseries.

Consequently, the objective of minimizing the global loss function is achieved by minimizing the weighted least squares expression in Equation~\eqref{eq:loss}. The global estimator takes the following form
\begin{align}
\label{eq:dlsa}
\widetilde{\theta} = \left(\sum_{k=1}^{K}T_k\widehat{\Sigma}_{k}^{-1}\right)^{-1}\left(\sum_{k=1}^{K} T_k \widehat{\Sigma}_{k}^{-1}\widehat{\theta}_{k}\right).
\end{align}
Then the covariance matrix of the estimated global parameters is given by
$\widetilde{\Sigma} = T\left(\sum_{k=1}^{K} T_k \widehat{\Sigma}_{k}^{-1}\right)^{-1}$.

Note that instead of simply averaging the estimated parameters, the analytical solution in
Equation~\eqref{eq:dlsa} approximates global estimators by taking a weighted average of
local parameters using weights $\widehat{\Sigma}_{k}^{-1}$, eliminating the influence of
outliers on the forecast performance when there are subseries that are poorly fitted by
ARIMA models. Furthermore, the simple averaging method assumes subseries as homogeneous
from worker to worker to guarantee statistical efficiency. This is highly questionable in
practice because heteroskedasticity and imbalance are common phenomenons for distributed
stored data.

The analytical form of the global estimator in Equation~\eqref{eq:dlsa} can be used to
estimate the global parameters in distributed forecasting. The difficulty with this method
is that it requires knowledge of $\widehat{\Sigma}_{k}$. Specifically, the local
parameters of a subseries in Equation~\eqref{eq:ar} are derived from a seasonal ARIMA
model and it may not be possible to straightly obtain a good estimate of the covariance
matrix. Moreover, a large order $p^{*}$ is often considered for a better approximation of
the initial ARIMA model trained by the split subseries, resulting in a set of local
parameters in which some entries are close to or equal to zero and a high-dimensional
covariance matrix whose dimension $p^{*}$ is likely to be larger than the length of
subseries. Therefore, following the literature of high-dimensional covariance matrix
estimation \citep{fan2008high,fan2011high,hyndman2011optimal}, we assume
$\widehat{\Sigma}_{k}$ is sparse and approximate it using $\hat{\sigma}_{k}^{2}I$ for each
subseries in this study, which greatly simplifies the computations and further reduces the
communication costs in distributed systems.

\subsection{Point Forecasts}
\label{sec:point}

After combining the local estimators from each subseries by minimizing the global loss function, the coefficients of the global estimators are calculated as illustrated in Section~\ref{sec:dlsa}. By using a large order $p^{*}$ instead of the infinite order in the converted AR representation for each subseries, the combined global model can be written generally as follows:
\begin{align}
\label{eq:forec}
y_t = \tilde{\beta}_0 + \tilde{\beta}_1 t + \sum_{i=1}^{p^{*}}\tilde{\pi}_{i}y_{t-i} + \sum_{j=1}^{l}\tilde{\eta}_{j}\bigg(\gamma_{j,t}-\sum_{i=1}^{p^{*}}\tilde{\pi}_{i}\gamma_{j,t-i}\bigg) + e_{t},
\end{align}
where $\tilde{\theta} = (\tilde{\beta}_0, \tilde{\beta}_1, \tilde{\pi}_1, \cdots , \tilde{\pi}_{p^{*}}, \tilde{\eta}_{1}, \cdots, \tilde{\eta}_{l})^{\top}$ is a vector of global model coefficients, and $e_{t}$ is the observed residual.

Given the time series $\{y_{t}\}$, suppose that we are at the time $T$ and are interested in forecasting the next $H$ observations, where the time index $T$ is the forecast origin. The $h$-step-ahead forecast can be calculated with relative ease as
\begin{align*}
\hat{y}_{T+h|T} =& \tilde{\beta}_{0}+\tilde{\beta}_{1} (T+h) + \sum_{j=1}^{l}\tilde{\eta}_{j}\bigg(\gamma_{j,T+h}-\sum_{i=1}^{p^{*}}\tilde{\pi}_{i}\gamma_{j,T+h-i}\bigg) \\
& +
  \begin{cases}
    \sum_{i=1}^{p^{*}}\tilde{\pi}_{i} y_{T+1-i}, & h=1 \\
    \sum_{i=1}^{h-1}\tilde{\pi}_{i} \hat{y}_{T+h-i|T} + \sum_{i=h}^{p^{*}}\tilde{\pi}_{i} y_{T+h-i}, & 1< h < p^{*} \\
    \sum_{i=1}^{p^{*}}\tilde{\pi}_{i} \hat{y}_{T+h-i|T}, & h \geq p^{*}
  \end{cases}.
\end{align*}
In this way, the point forecasts of the next $H$ observations can be calculated recursively for $h=1,\dots,H$.

\subsection{Prediction Intervals}
\label{sec:pis}

As described in Section~\ref{sec:arima2ar}, the linear representation of the seasonal ARIMA model, which is trained for each subseries of $\{y_{t}\}$, is derived from the following AR model for time series $\{x_{t}\}$:
\begin{align*}
x_t = \sum_{i=1}^{p^{*}} \pi_{i} x_{t-i}+\varepsilon_{t},
\end{align*}
where the infinite order for the converted AR model is replaced by a large order $p^{*}$. As
is well-known, once we estimate the coefficients of the AR regression and the standard
deviation of the residuals, the standard error of the $h$-step ahead forecast can be
uniquely determined. Thus, the forecast variances of the linear representation in
Equation~\eqref{eq:ar} are not affected by the intercept term and time trend term of the
seasonal ARIMA model (ignoring estimation error). Consequently, the forecast variances of
the combined global model in Equation~\eqref{eq:forec} depend only on the AR part of the
model, that is the term $\sum_{i=1}^{p^{*}}\tilde{\pi}_{i}y_{t-i} + e_{t}$.

To compute these variances, we convert the AR model to a MA model with infinite order
\citep{brockwell2016introduction}:
\begin{align*}
e_{t} + \sum_{i=1}^{\infty}\tilde{\psi}_{i}e_{t-i}.
\end{align*}
Then, in the global model, the standard error of the $h$-step ahead forecast is given by
\begin{align*}
\tilde{\sigma}_{h}^{2} = \begin{cases}
  \tilde{\sigma}^{2}, & h = 1 \\
  \tilde{\sigma}^{2} \left( 1 + \sum_{i=1}^{h-1}\tilde{\psi}_{i}^{2} \right), & h > 1,
\end{cases}
\end{align*}
where $\tilde{\sigma}$ is the standard deviation of the residuals for the combined global model and is unknown. As illustrated in Section~\ref{sec:dlsa}, we suggest replacing the covariance estimate $\widehat{\Sigma}_{k}$ of local estimators with $\hat{\sigma}_{k}^{2} I$. Subsequently, the covariance estimate of the global estimators is calculated by $\widetilde{\Sigma} = \left(\sum_{k=1}^{K}\left(T_{k} / T\right) (\hat{\sigma}_{k}^{2} I)^{-1}\right)^{-1}$, and we can estimate $\tilde{\sigma}^{2}= \operatorname{tr}(\widetilde{\Sigma})/p$.

Assuming normally distributed errors, the central $100(1-\alpha)\%$ prediction interval for the $h$-step ahead forecast is given by
\begin{align*}
\hat{y}_{T+h|T} \pm \Phi^{-1}(1-\alpha/2)\tilde{\sigma}_{h},
\end{align*}
where $\Phi$ is the cumulative distribution function of the standard normal distribution.

\section{Application to Electricity Data}
\label{sec:application}

In the section, the electricity demand data set we use to investigate the performance of
the proposed distributed ARIMA models and the experimental design are shown in detail.  We
analyze the performance of the proposed distributed ARIMA models and explore the factors
that affect its forecasting performance.

\subsection{Data Description}
\label{sec:data}

To illustrate our proposed approach, we forecast the time series of the GEFCom2017
\citep{hong2019global}. The data, publicly available at
\url{https://github.com/camroach87/gefcom2017data}, was initially made available by ISO
New England. It comprises the electricity load, holiday information, and weather data
composed of dew point and dry bulb temperatures. To assess the benefits of the proposed
distributed ARIMA model, we restrict our attention to the electricity load data in the
following analysis.

\begin{figure}[!ht]
    \centering
    \includegraphics[width=\textwidth]{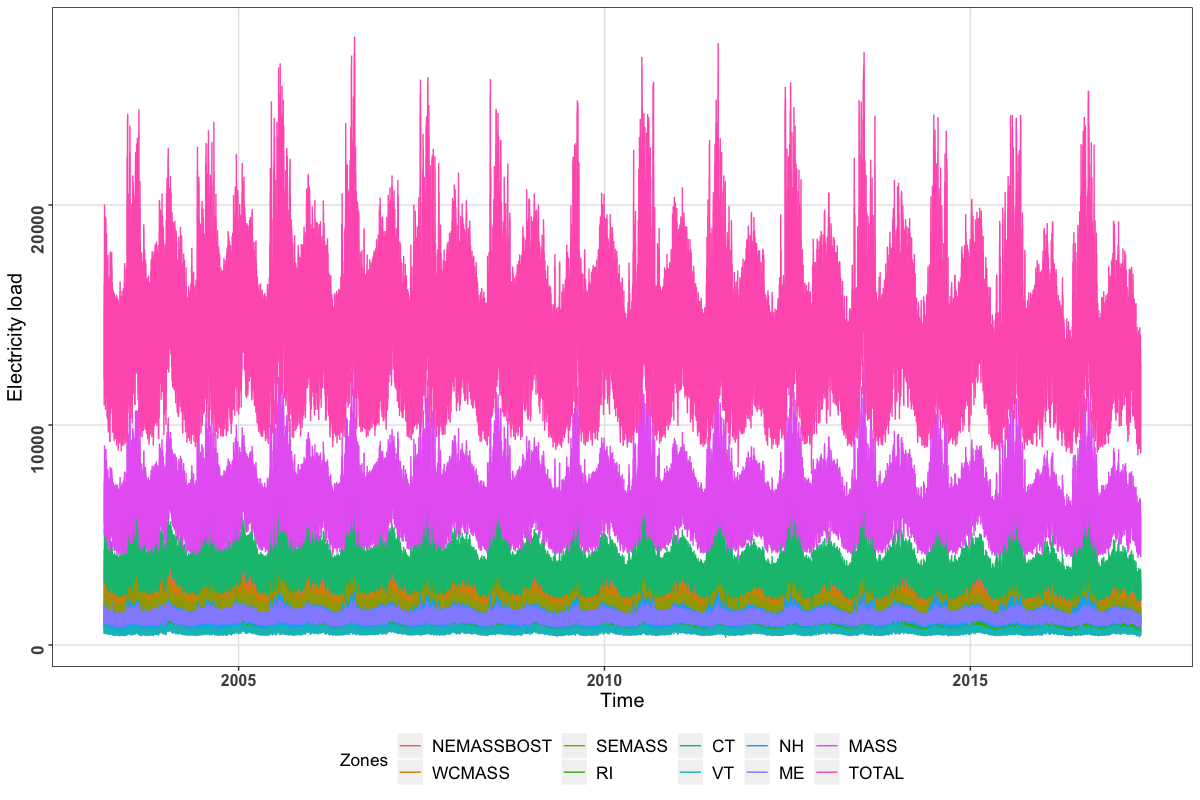}
    \caption{Time series plot of the electricity load data (in megawatt) for eight bottom-level zones and two aggregated zones.}
    \label{fig:tsplot}
\end{figure}

The electricity load data set consists of $10$ time series of hourly data, ranging from 1
March 2003 to 30 April 2017. As aforementioned in Section~\ref{sec:arima}, the
computational time for the automatic ARIMA modeling grows exponentially over the time
length, reaching $10$ minutes at the length of $1,000$. Allowing a wider range of model
orders will further increase the execution time beyond an acceptable range (see
Section~\ref{sec:factors} for more experimental results). In this respect, the electricity
load data, spanning $124,171$ time points, is long enough so that distributed computing is
desired, consistent with the applicable scenarios of our proposed approach.
Dealing with data of this magnitude is uncommon in traditional forecasting applications,
but it should be noted that load forecasting problems routinely have cases with available
data spanning more than one million time points, most likely for several locations at the same time.
The forecasts are primarily used in power systems operations, such as energy trading and unit commitment.

Figure~\ref{fig:tsplot} presents the hourly electricity load for all
bottom-level zones and two aggregated zones. We train the distributed ARIMA model using
data from 1 March 2003 to 31 December 2016, while data from 1 January 2017 to 30 April
2017 are used for testing. In this way, we provide the four-month ($2879$-step) ahead
point forecasts, and the corresponding prediction intervals with multiple confidence
levels. The original GEFCom2017 \citep{hong2019global} only requires one month ahead
forecasting. However, forecasting at longer horizons in energy and many other
high-frequency time series domains is in great demand as it allows for earlier management
plans. Note that we restrict our attention to time series forecasting on distributed
systems without considering the data's hierarchical configuration.

Electricity demand may exhibit periodic patterns such as time of the day, day of the week,
and month of the year. Although day of the week can be easily involved in our proposed
model as covariates, our ex-ante analysis shows no significant distinction in patterns
between days of the week: whether the day-of-the-week pattern is included in ARIMA models
or distributed ARIMA models results in little difference in both point and interval
forecasting accuracy. Moreover, the subseries is not long enough to enable us to consider
monthly seasonality under distributed computing environment, while the monthly seasonality
can be handled with preprocessing steps such as time series decomposition. To better focus
on assessing the benefits of the proposed distributed ARIMA models over normal ARIMA
models, we only consider the time-of-the-day effect in the following analysis by using the seasonal
components of ARIMA models for hourly subseries ($m=24$).

\subsection{Experimental Setup}
\label{sec:setup}

We partition each time series into $150$ subseries in the experiment with the length of
each subseries about $800$. The setup is inspired by the M4 competition
\citep{makridakis2020m4}: the length of the hourly time series ranges from $700$ to
$900$. For time series with such lengths, traditional forecasting models perform well on a
standalone computer, and the time consumed by automatic ARIMA modeling process is within
$5$ minutes, which is empirically acceptable. The analysis exploring the forecasting
performance on different settings of the number of subseries is presented in
Section~\ref{sec:factors}.

As illustrated in Section~\ref{sec:arima2ar}, the AR representation with infinite order is
obtained from the seasonal ARIMA model for each subseries to facilitate the
parameter-based combination. We approximate the infinite order AR model with one sizeable
finite order, balancing model complexity, and approximating the original ARIMA model.
In practice, one can perform an ex-ante analysis using several candidate order values and
select the smallest one from the order values resulting in no significant differences
in the forecasting results of the original ARIMA model and its converted linear
representation. Alternatively, if no ex-ante analysis has been conducted, a sizeable
order is always recommended as it will only lead to negligible computational
complexity in the combination of local parameters in the master node, while allowing a
broader range of order values for the original ARIMA models. Accordingly, we set the AR order
$p^{*}$ to $2000$ in this experiment, see Section~\ref{sec:results} for more details.

In this paper, the distributed ARIMA model is designed with the aim of facilitating the ARIMA
modeling for ultra-long time series in a distributed manner with high computational efficiency
and potentially improved forecasting performance, rather than running a horse race between the
proposed method and other forecasting methods. Therefore, the main comparison we are interested
in is DARIMA versus ARIMA. Moreover, the most widely used forecasting
approaches developed with distributed systems are poorly scalable for large forecast horizons,
making it impractical to apply these approaches to forecasting quite a few future values
(see Section~\ref{sec:tsforecast} for more details). In
this regard, we only compare the proposed approach to ARIMA models for the whole time series,
as well as an additional standard for comparison: ETS models.

For comparison purposes, the argument configuration of the automatic ARIMA modeling for
the whole series and subseries is shown in Table~\ref{tab:config}. To make the algorithm
execution time comparable, we consider the global order selection with parallelism in
fitting models for the whole time series, while using non-parallel stepwise order
selection when modeling the subseries. Furthermore, we apply the CSS method to fit ARIMA
models instead of CSS-ML (see Section~\ref{sec:arima} for details). With the fitting
method CSS-ML, we may have to fit more than one model in the model refit process, since
the model with the appropriate order identified in the order selection process may be
rejected by several strict checks for unit roots. Due to the uncertainty, the comparison
of execution time between the ARIMA model on the whole series and the distributed ARIMA
model would be unreliable if we used CSS-ML\@. Finally, the experiment is limited to
specific maximum values of model orders. We further discuss the importance of model orders
to forecasting performance in Section~\ref{sec:factors}. Moreover, implementations of ETS
models are available in the \pkg{forecast} \proglang{R} package with the \code{ets()} function.
Given a single time series data, ETS modeling can not be extended to parallel computing.

\begin{table}
\centering
\caption{Argument configuration of the automatic ARIMA modeling for the whole series and
  subseries respectively, where ARIMA denotes the automatic ARIMA model for the whole time
  series and DARIMA denotes the distributed ARIMA model. The argument \texttt{max.order}
  represents the maximum value of $p+q+P+Q$ in the process of global order selection. ARIMA
  models are implemented using the \code{auto.arima()} function in the \pkg{forecast} package
  for \proglang{R}.}
\begin{tabular}{lcc}
\toprule
  Argument                       & ARIMA & DARIMA \\
  \midrule
  \texttt{max.p}; \texttt{max.q} & 5     & 5      \\
  \texttt{max.P}; \texttt{max.Q} & 2     & 2      \\
  \texttt{max.order}             & 5     & 5      \\
  \texttt{fitting method}        & CSS   & CSS    \\
  \texttt{parallel} (multicore)  & True  & False  \\
  \texttt{stepwise}              & False & True   \\
\bottomrule
\end{tabular}
\label{tab:config}
\end{table}

As for the system environment, the experiments are carried out on a Spark-on-YARN cluster on
Alibaba Cloud Server composed of one master node and two worker nodes. Each node contains
$32$ logical cores, $64$ GB RAM and two $80$ GB SSD local hard drives. The algorithm is
developed on Spark platform (2.4.5), and both \proglang{Python} as well as \proglang{R}
interfaces are freely available at \url{https://github.com/xqnwang/darima}.

\subsection{Evaluation Measures}
\label{sec:eval}

To assess the performance of the point forecasts, we consider the mean absolute scaled
error \citep[MASE;][]{hyndman2006another} as the measure of forecasting accuracy. MASE is
recommended because of its excellent mathematical properties, such as scale-independent
and less insensitive to outliers. Besides, \citet{hyndman2006another} suggest MASE as the
standard measure for comparing forecasting accuracy across multiple time series. The formula
for computing the MASE is the following:
\begin{align*}
\text{MASE}&=\frac{\frac{1}{H}\sum\limits_{t=T+1}^{T+H}|y_{t}-\hat{y}_{t|T}|}{\frac{1}{T-m}\sum\limits_{t=m+1}^{T}|y_{t}-y_{t-m}|}.
\end{align*}

We evaluate the accuracy of prediction intervals using the mean scaled interval score
\citep[MSIS;][]{gneiting2007strictly}, given by
\begin{align*}
\text{MSIS} = \frac{\displaystyle
    \frac{1}{H}\sum_{t=T+1}^{T+H} (U_{t|T}-L_{t|T}) +
    \frac{2}{\alpha} (L_{t|T}-y_{t}) \bm{1}\{y_{t}<L_{t|T}\} +
    \frac{2}{\alpha} (y_{t}-U_{t|T}) \bm{1}\{y_{t}>U_{t|T}\}
  }
  {\displaystyle
    \frac{1}{T-m} \sum_{t=m+1}^{T}|y_{t}-y_{t-m}|
  },
\end{align*}
where $L_{t|T}$ and $U_{t|T}$ are lower and upper bounds of the generated
$100(1-\alpha)$\% prediction interval, respectively. The scoring rule balances the width of
the generated prediction intervals and the penalty for true values lying outside the
prediction intervals.

\subsection{Distributed Forecasting Results}
\label{sec:results}

We now investigate the performance of the proposed distributed ARIMA models on the
GEFCom2017 data set compared to that of ARIMA models as well as ETS models in terms
of MASE as well as MSIS\@. Execution time is
also considered as an important metric describing the computational efficiency of
algorithms. For conciseness, our proposed algorithm, the \emph{distributed} ARIMA model, is
hereinafter referred to as DARIMA.

To verify whether the approximating order of the AR representation of $2000$ (as described
in Section~\ref{sec:setup}) is large enough to make the AR model close to its original
seasonal ARIMA model, we present the forecasting results of the ARIMA model and its
AR$(2000)$ representation on the GEFCom2017 data set in Table~\ref{tab:score}. We observe
that there is no difference between the forecasting performance of the ARIMA model and
that of the converted AR model (as measured by MASE and MSIS), to the degree of $10^{-3}$.

\begin{table}
  \centering
  \caption{Benchmarking the performance of DARIMA against ARIMA models and their AR
    representations, as well as ETS models with regard to MASE and MSIS\@.
    For each measure, the minimum score among the four algorithms is marked in \textbf{bold}.}
  \begin{tabular}{lcccccc}
    \toprule
                      & \multicolumn{3}{c}{MASE} & \multicolumn{3}{c}{MSIS}                           \\
    \cmidrule(lr){2-4} \cmidrule(lr){5-7}
                      & Mean                     & Median         & SD              & Mean            & Median          & SD             \\
    \midrule
    DARIMA            & \textbf{1.297}           & \textbf{1.218} & \textbf{0.284}  & \textbf{15.078} & \textbf{14.956} & \textbf{1.021} \\
    ARIMA             & 1.430                    & 1.325          & 0.351           & 19.733          & 16.498          & 7.446          \\
    AR representation & 1.430                    & 1.325          & 0.351           & 19.733          & 16.498          & 7.446          \\
    ETS               & 1.491                    & 1.338          & 0.408           & 53.783          & 49.109          & 15.834         \\
    \bottomrule
  \end{tabular}
  \label{tab:score}
\end{table}

Table~\ref{tab:score} also compares the forecasting performance of DARIMA against ARIMA and ETS
for the whole time series in terms of the mean, median, and standard deviation (SD) of the MASE and MSIS values. As
expected, DARIMA always outperforms the benchmark methods regardless of point forecasts or
prediction intervals. Specifically, for point forecasting, DARIMA achieves substantial
performance improvements compared to the benchmark methods, approximately at least $9.3\%$ for the
mean MASE value and $8.1\%$ for the median, with a smaller degree of variation. Meanwhile, DARIMA yields a statistically
significant improvement (at least $23.6\%$) over the benchmark methods in terms of the mean of
MSIS values. Therefore, implementing ARIMA models on distributed systems by splitting the
whole time series into several subseries dominates ARIMA and ETS in both
point forecasting and interval forecasting.

\begin{figure}
    \centering
    \includegraphics[width=\textwidth]{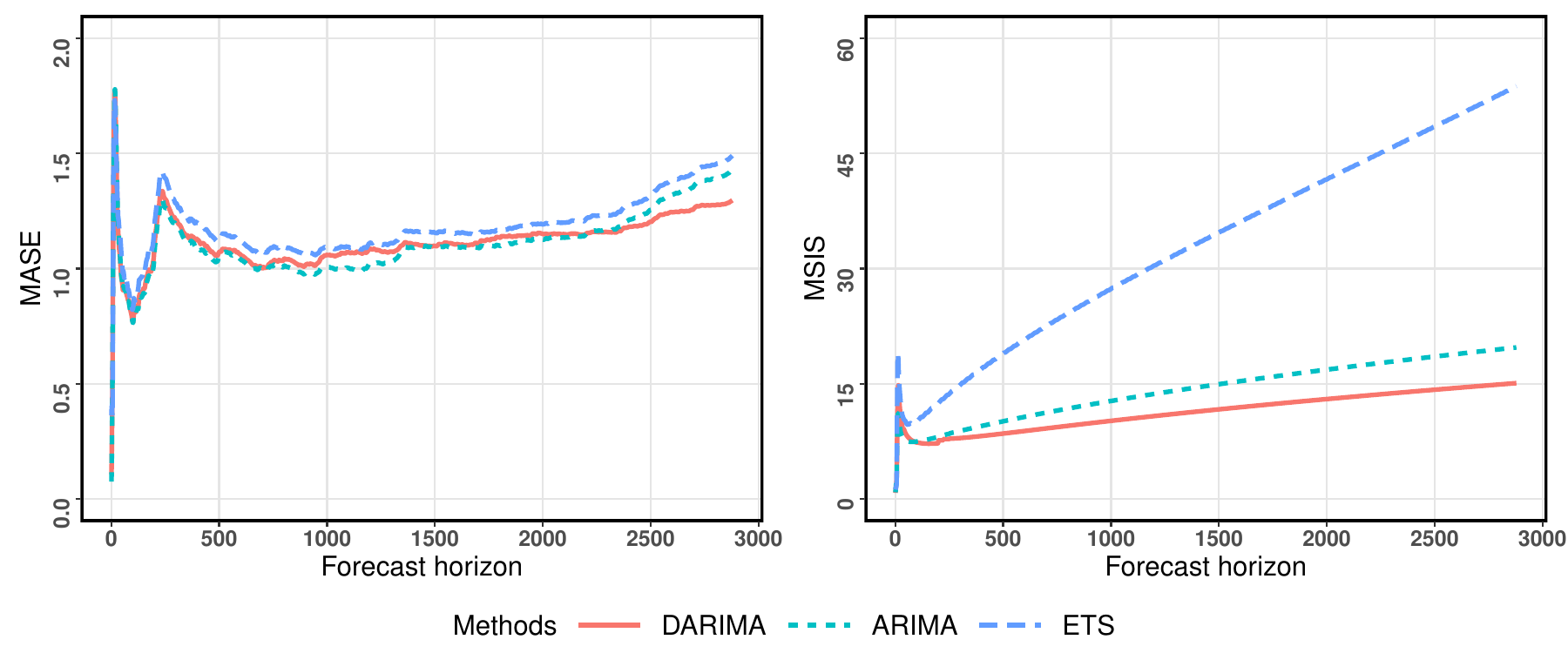}
    \caption{Benchmarking the performance of DARIMA against ARIMA and ETS for various forecast
      horizons. Comparisons are presented regarding the mean of MASE as well as MSIS
      values.}
    \label{fig:mcumscore}
\end{figure}

We proceed by observing how the forecasting performance of distributed ARIMA models
changes with the forecast horizon. Figure~\ref{fig:mcumscore} depicts the accuracy of
DARIMA over various forecast horizons against the benchmark methods: ARIMA and ETS\@. First, the
left panel shows that the point forecasting performance of DARIMA displays small
differences with ARIMA and ETS when we are interested in obtaining the forecasts of the first few
future values. We also observe that DARIMA yields slightly larger values than ARIMA in
terms of MASE when focusing on forecasting the next $1000$ observations. This difference
tapers off with increasing forecast horizon, and finally, DARIMA significantly outperforms
both ARIMA and ETS for the forecasting of long-term observations. On the other hand, the right panel
illustrates that DARIMA provides much better performance than ARIMA and ETS according to MSIS
values when we turn our attention to forecasting more than $100$ future
values. The achieved performance improvements become more pronounced as the
forecast horizon increases. Furthermore, one possible reason for the result that
DARIMA and ARIMA models perform substantially better than ETS models
is that, for seasonal data, there are many more ARIMA models than the possible
models in the exponential smoothing class \citep{Hyndman2008b}.
In simple terms, we conclude that if long-term observations are
considered, DARIMA is favorable, both in point forecasts and prediction intervals.

\begin{figure}
    \centering
    \includegraphics[width=0.85\textwidth]{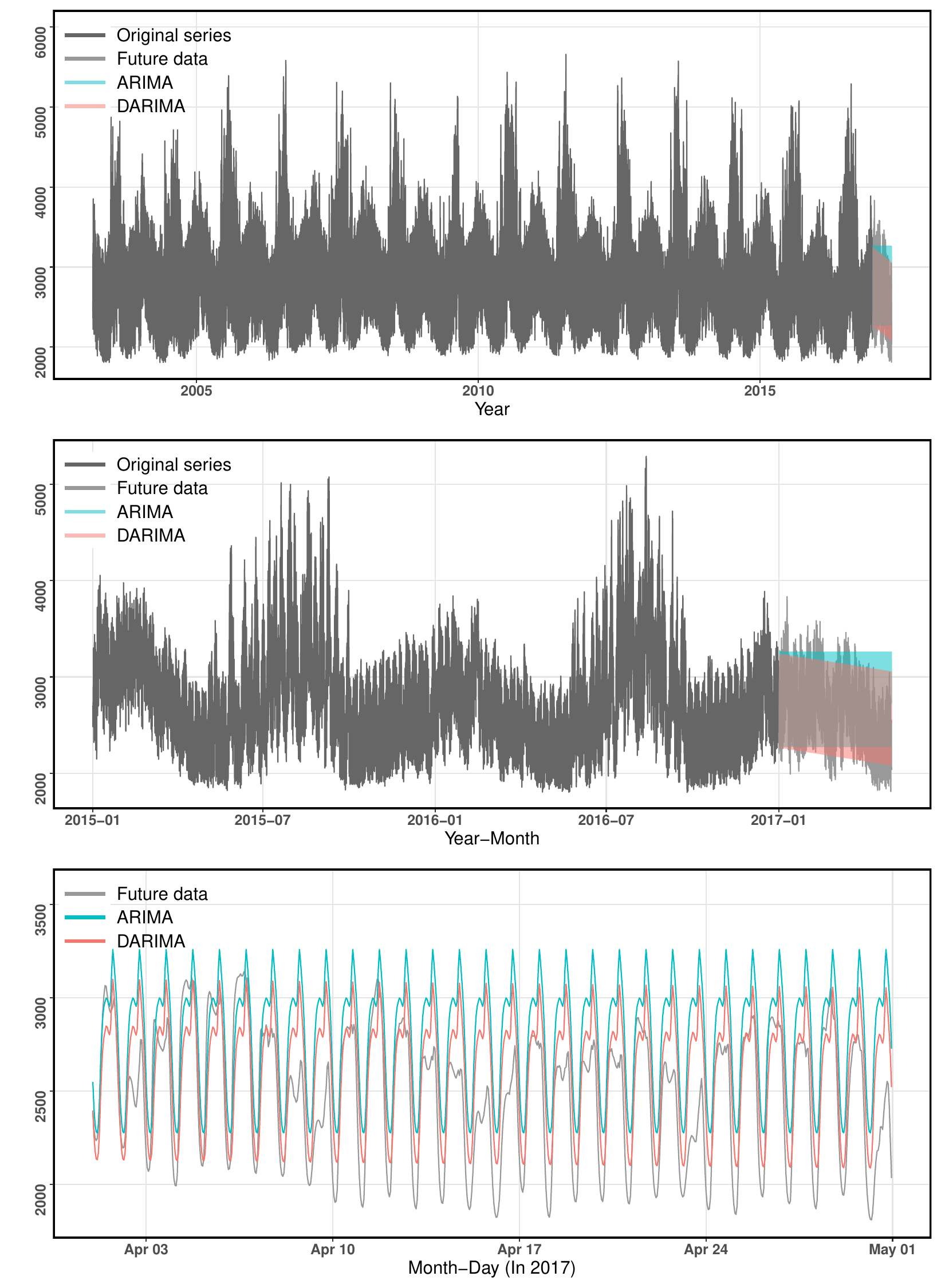}
    \caption{An example showing the electricity demand series for the NEMASSBOST zone, and
      forecasts using the proposed approach and the benchmark method, ARIMA on different
      zoom levels. The top panel depicts the original series, future data, as well as the
      forecasts of ARIMA and DARIMA\@. The middle panel shows a clip from 1 January 2005
      to 30 April 2017, while the bottom panel shows a shorter clip of April 2017 to
      illustrate the forecasting performance on the large forecast horizon.}
    \label{fig:forecastplot}
\end{figure}

Figure~\ref{fig:forecastplot} shows the forecasting performance of DARIMA compared to
ARIMA on the electricity demand series for the NEMASSBOST zone. We observe from the
forecasts that, compared to ARIMA, DARIMA captures the yearly seasonal pattern from the
original series. Even for large forecast horizons, DARIMA results in forecasts
closer to the true future values than ARIMA\@. These conclusions are consistent with the
previous results shown in Table~\ref{tab:score} and Figure~\ref{fig:mcumscore}.

\begin{figure}
    \centering
    \includegraphics[width=0.75\textwidth]{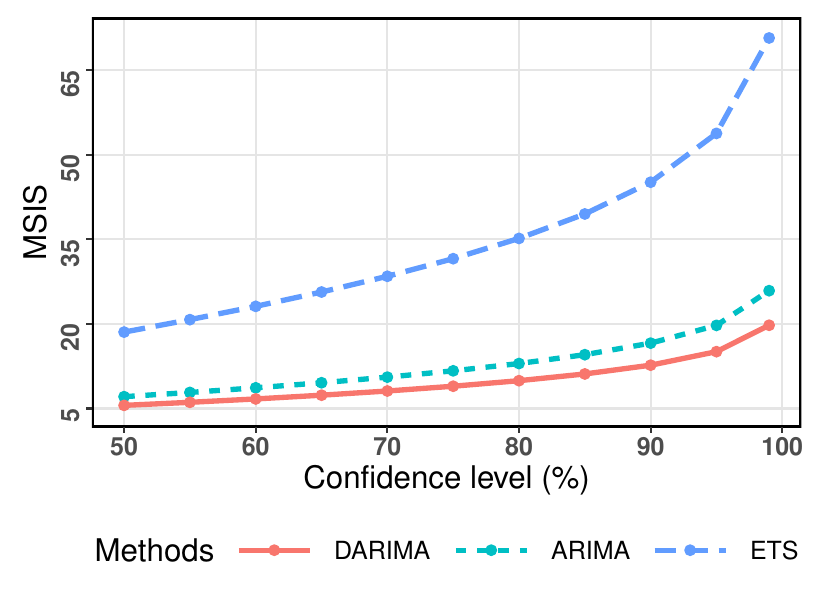}
    \caption{The comparison between the forecasting performance of DARIMA,
      ARIMA, and ETS in terms of MSIS across different confidence levels.}
    \label{fig:msis}
\end{figure}

Figure~\ref{fig:msis} presents the MSIS results of forecasting with DARIMA,
ARIMA, and ETS across different confidence levels varying from $50\%$ to $99\%$. We observe that
DARIMA persistently results in better forecasting accuracy than ARIMA and ETS in terms of MSIS
across various confidence levels. Besides, the superiority of DARIMA tends to be more
substantial as the confidence level increases.

\begin{figure}[!ht]
    \centering
    \includegraphics[width=0.75\textwidth]{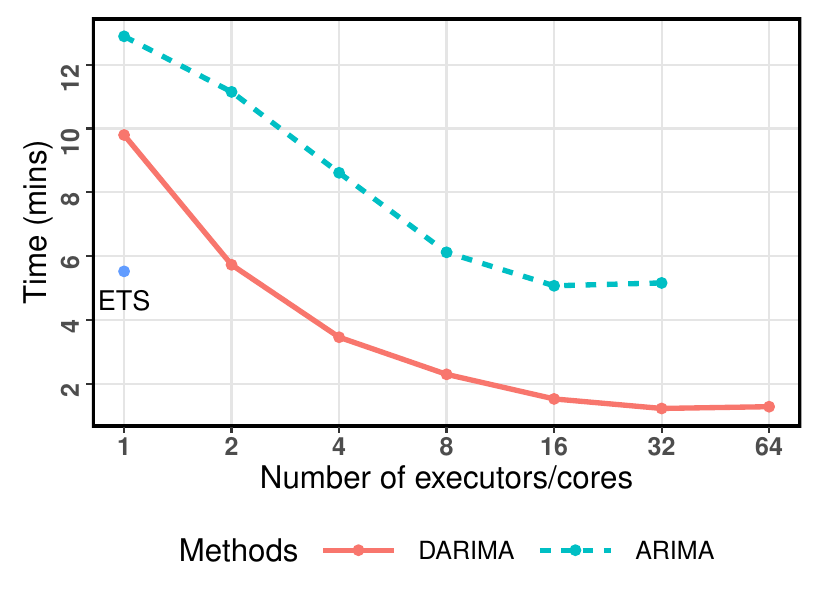}
    \caption{The comparison between the average execution time of DARIMA, ARIMA, and
    ETS modeling for a single time series on the GEFCom2017 data set with different
    numbers of executors/cores. Limited by the hardware of the device
    (each node contains only 32 virtual cores), the execution time of the ARIMA on 64
    cores is not available.}
    \label{fig:time}
\end{figure}

The aforementioned results mainly focus on the forecasting accuracy of DARIMA against the
benchmark methods. Now we compare DARIMA to ARIMA and ETS in terms of execution time to
investigate the computational efficiency of DARIMA, as shown in Figure~\ref{fig:time}.
The forecasting performance of DARIMA varies with the number of subseries (see
Section~\ref{sec:factors} for details), but not with the number of executors
in the Spark system. More specifically, the number of executors determines how many tasks
(\ie~subproblems of ARIMA modeling) are assigned to each executor, which greatly affects
the runtime in individual executors. When it comes to ARIMA
modeling for the whole time series, increasing the number of cores used can also give a significant
speedup as the specification search is done in parallel. However, for a given ultra-long
time series, ETS modeling can not be implemented in a parallel manner.
So the execution time of ETS modeling does not change along with the number of virtual cores used.
Figure~\ref{fig:time} shows improved computational efficiency of both ARIMA and DARIMA with increasing
numbers of executors/cores. Besides, DARIMA persistently results in less execution time
than ARIMA and ETS when using more than two executors/cores. In our application, modeling a DARIMA
model for an ultra-long time series with the length of about $120,000$ takes an average of
$1.22$ minutes with $32$ cores, while ARIMA modeling takes an average of $5.16$
minutes and ETS takes an average of $5.38$ minutes. Therefore, our approach results in significantly
improved forecasting accuracy with remarkably less execution time compared to ARIMA and ETS models.

\subsection{Sensitivity Analysis}
\label{sec:factors}

This subsection focuses on the factors that may affect the forecast performance of the
distributed ARIMA models. In the following analysis, we consider two main factors: the
number of split subseries and the maximum values of model orders. Other potential factors
will be discussed in Section~\ref{sec:discussion}.

\begin{figure}
    \centering
    \includegraphics[width=\textwidth]{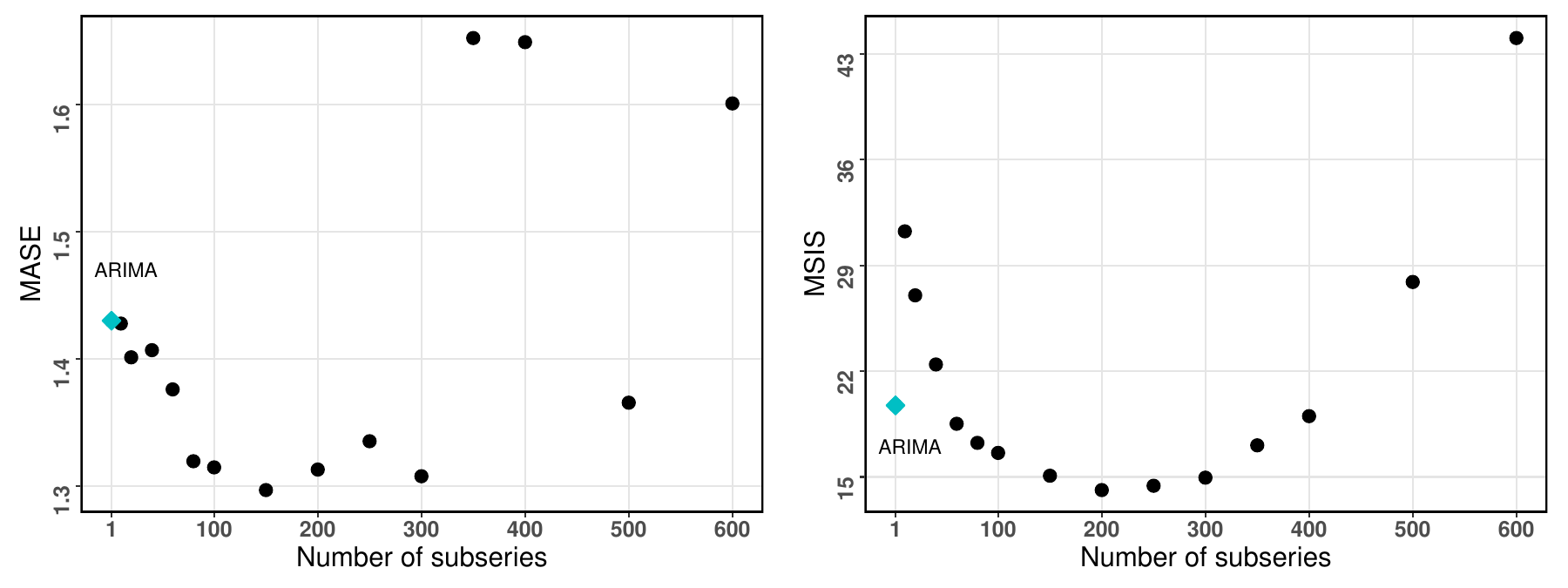}
    \caption{Relationship between the forecasting performance of the distributed ARIMA
      models and the number of subseries on the GEFCom2017 data set. The score of ARIMA for
      the whole series which equals to that of the distributed ARIMA models with only one
      subseries, is shown in the diamond point.}
    \label{fig:subseries}
\end{figure}

We first explore the relationship between the forecasting performance of the distributed
ARIMA models and the number of subseries $K$ preset in the preprocessing process, as
presented in Figure~\ref{fig:subseries}. In essence, the relationship also depicts the
importance of the length of subseries to the functioning of the distributed ARIMA
models. With the number of subseries $K$ varying from $10$ to $100$, there is a
considerable drop in the MASE values of DARIMA\@. It then slightly fluctuates when $K$ is
between $100$ and $300$, and has an enormous growth when $K$ equals to
$350$. Subsequently, the MASE values of DARIMA go up and down widely with a larger
$K$. Besides, the MSIS of DARIMA shows an overall trend of decreasing first and then
increasing. Therefore, we conclude that the number of subseries should be controlled
within a reasonable range, with too few or too many subseries causing poor forecasting
performance. In our study, we should limit the number of subseries between $100$ to $300$.

Table~\ref{tab:maxorder} compares the forecasting performance of DARIMA with that of ARIMA
under different settings of the maximum values of model orders in terms of MASE and
MSIS\@. The maximum value of $p+q+P+Q$ only works for the process of global order
selection. Therefore, when we keep the maximum values of non-seasonal and seasonal parts
fixed, the changes in the maximum value of $p+q+P+Q$ result in some changes in the
forecasting accuracy of ARIMA, but no changes in that of the DARIMA\@. If the model orders
are allowed to range more widely, ARIMA achieves better forecasting performance on both
point forecasts and prediction intervals. The main reason is that the broader range of
model orders provides more possible models in the order selection process. In contrast,
DARIMA performs higher MASE when more possible models are provided. One possible reason
for this result is that allowing more extensive choices of model orders may lead to overfitting
for subseries with short lengths. Moreover, Table~\ref{tab:maxorder} shows that the
maximum values of model orders have a limited impact on forecasting performance: the
changes in performance both for ARIMA and DARIMA gradually disappear as the maximum orders
increases. We also compare the results using the symmetric mean absolute percentage error
\citep[sMAPE;][]{makridakis1993accuracy} and obtain almost identical results from those
with MASE\@. As expected, DARIMA always outperforms ARIMA on different settings of the
maximum values of model orders for both point forecasts and prediction intervals.

We proceed by comparing our proposed DARIMA with ARIMA regarding execution time on
different settings of the maximum values of model orders, as shown in
Table~\ref{tab:maxorder}. The results indicate that DARIMA is more computationally
efficient than ARIMA in multiple settings of the maximum values of model orders. When the
model orders are allowed to take a broader range of values, both DARIMA and ARIMA take
more time modeling the time series. The execution time spent on ARIMA modeling has a
marked increase, while DARIMA keeps its modeling time within a reasonable and acceptable
range. For example, DARIMA is $53$ times more efficient than ARIMA on the setting of
max.orders being equal to $(8,4,10)$. The improved efficiency makes it possible for DARIMA to search for an
appropriate model for each subseries in a broader range of candidate models.

\begin{table}[!ht]
\centering
\caption{Performance comparison of DARIMA and ARIMA on different settings of the maximum
  values of model orders in terms of MASE, MSIS as well as execution time over $30$
  executors/cores. The argument max.orders in the first column, is composed of three
  components: the maximum value of $p$ (equals to that of $q$), the maximum value of $P$
  (equals to that of $Q$) and the maximum value of $p+q+P+Q$. For each measure, the lowest
  value of the scoring rule under a specific order setting is presented in \textbf{bold}.}
\renewcommand{\arraystretch}{0.8}
\begin{tabular}{clrrr}
\toprule
Max orders & Method & MASE           & MSIS            & Execution time \\ [-0.3cm]
           &        &                &                 & (mins)         \\
\midrule
(5, 2, 5)  & ARIMA  & 1.430          & 19.733          & 4.596          \\
           & DARIMA & \textbf{1.297} & \textbf{15.078} & \textbf{1.219} \\ [0.2cm]
(5, 2, 7)  & ARIMA  & 1.410          & 18.695          & 14.189         \\
           & DARIMA & \textbf{1.297} & \textbf{15.078} & \textbf{1.211} \\ [0.2cm]
(6, 2, 7)  & ARIMA  & 1.410          & 18.695          & 15.081         \\
           & DARIMA & \textbf{1.298} & \textbf{15.108} & \textbf{1.326} \\ [0.2cm]
(6, 3, 7)  & ARIMA  & 1.413          & 15.444          & 21.072         \\
           & DARIMA & \textbf{1.324} & \textbf{12.590} & \textbf{1.709} \\ [0.2cm]
(6, 3, 10) & ARIMA  & 1.413          & 15.654          & 76.272         \\
           & DARIMA & \textbf{1.324} & \textbf{12.590} & \textbf{1.769} \\ [0.2cm]
(7, 3, 10) & ARIMA  & 1.413          & 15.654          & 83.077         \\
           & DARIMA & \textbf{1.327} & \textbf{12.561} & \textbf{1.829} \\ [0.2cm]
(7, 4, 10) & ARIMA  & 1.409          & 13.667          & 111.292        \\
           & DARIMA & \textbf{1.338} & \textbf{12.079} & \textbf{2.267} \\ [0.2cm]
(8, 4, 10) & ARIMA  & 1.409          & 13.667          & 117.875        \\
           & DARIMA & \textbf{1.335} & \textbf{12.076} & \textbf{2.224} \\
\bottomrule
\end{tabular}
\label{tab:maxorder}
\end{table}

\section{Numerical Simulations}
\label{sec:sim}

In this section, we perform a simulation study to further investigate and justify our proposed
DARIMA method in terms of forecasting accuracy and computational cost.

\subsection{Simulation Setup}
\label{sec:sim_setup}

We consider daily, hourly, and half-hourly series and generate $1,000$ time
series samples in each case. Each series is generated by an ARIMA$(p, d, q)(P, D, Q)_m$
process with $d$ being randomly sampled from Bernoulli($0.9$), $D$ being randomly sampled
from Bernoulli($0.4$), $p$ and $q$ each taking values from a uniform distribution on
$\{0,1,2,3,4,5\}$, and $P$ and $Q$ each taking values $0$, $1$ and $2$ with equal
probability. The periods $m$ of the simulated series are set to be $7$, $24$ and $48$ to
match daily, hourly and half-hourly time series. For each generated series, the
parameters of each process are randomly chosen from the uniform distribution U($-2,2$)
over the stationary and invertible space.

We divide each series into three parts: the first $m \times 10$ observations are discarded as
burn-in, the following $T$ observations are used as a training set for estimating parameters, and
the last $H$ observations are used for testing. For daily, hourly, and half-hourly series,
$T$ takes the values $8,000$, $100,000$ and $200,000$ respectively, while $H$ takes the
values $100$, $2,000$ and $4,000$ respectively.

Final forecasts are produced using four different methods:
(i) distributed ARIMA modeling as introduced in Section~\ref{sec:method} (DARIMA);
(ii) simply averaging the estimated parameters for split subseries when implementing DARIMA (DARIMA\_SA);
(iii) automatic ARIMA modeling for the whole series with a single model (ARIMA);
and (iv) ETS modeling for the whole time series (ETS).
For comparison purposes, DARIMA and DARIMA\_SA share the same
settings for time series partitions. Specifically, each daily series is partitioned into
$21$ subseries with each subseries spanning at least $52$ weeks, and each hourly and
half-hourly series is split into $138$ subseries so that each subseries spans at least
$30$ days.  Other settings are consistent with that in the application to electricity
data, see Section~\ref{sec:setup} for more details.

\subsection{Results}
\label{sec:sim_results}

We choose three metrics to evaluate the performance of our proposed method, which are MASE,
MSIS, and ACD (absolute coverage difference). To assess prediction intervals, we set
$\alpha = 0.05$ (corresponding to $95\%$ prediction intervals). As a
supplemental scoring rule, ACD measures the absolute difference between the actual coverage
of the target method and the nominal coverage, where the coverage measures the rate at which the
true values lie inside the prediction intervals the method provides.

\begin{figure}[!ht]
    \centering
    \includegraphics[width=\textwidth]{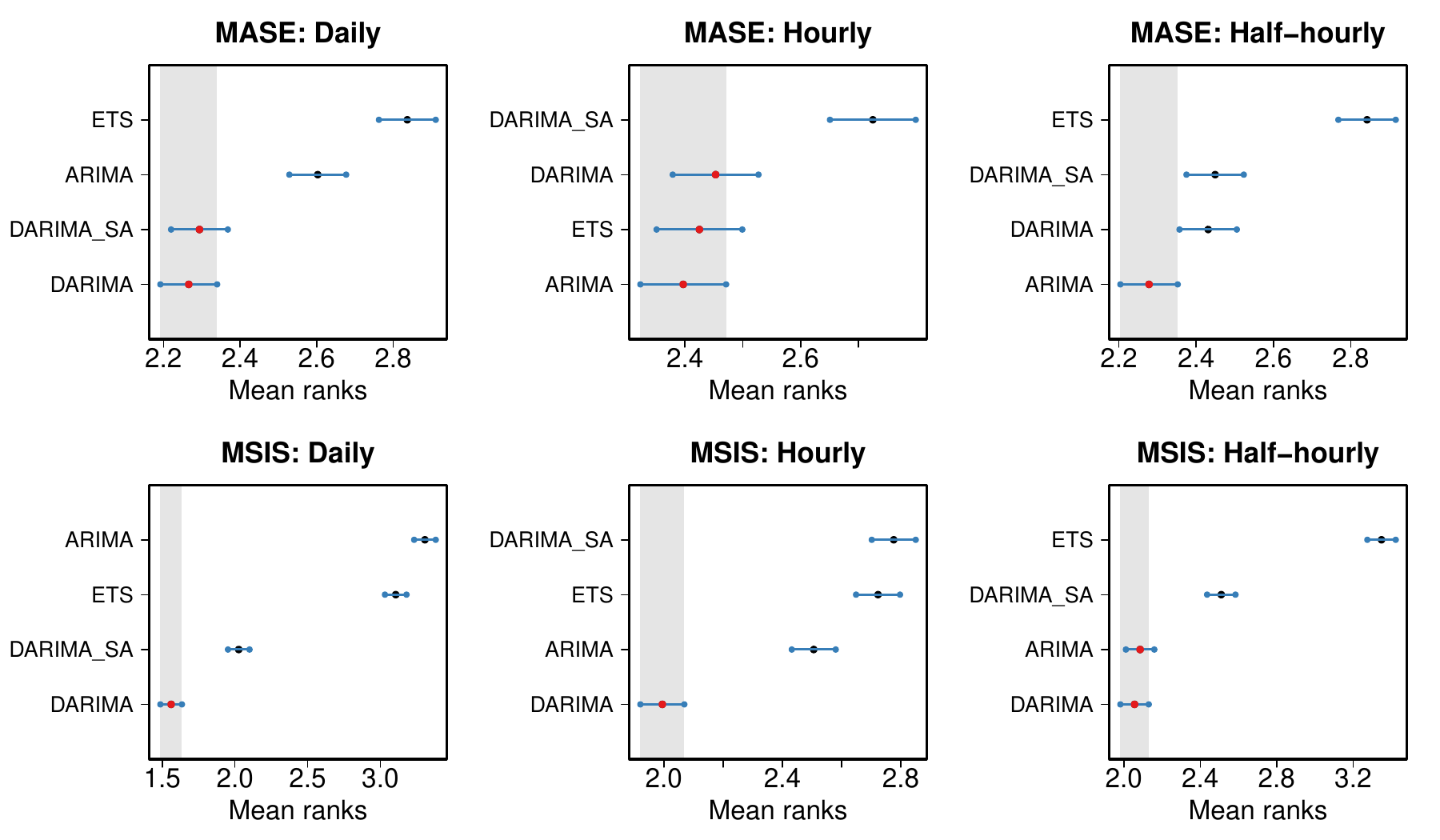}
    \caption{MCB significance tests for DARIMA, DARIMA\_SA, ARIMA, and ETS for each data frequency.}
    \label{fig:sim}
\end{figure}

Table~\ref{tab:sim} displays the forecasting accuracy of DARIMA as well as three methods
considered as benchmarks in this study. The accuracy is reported for short-term (four weeks)
and long-term (remaining periods) horizons separately as well as across all forecast horizons.
Moreover, the multiple comparisons with the best \citep[MCB,][]{koning2005m3} test is
performed on each data frequency to identify whether the average ranks, based on MASE and
MSIS, of the four models considered are significantly different, as presented in Figure~\ref{fig:sim}.
If the intervals of two methods do not overlap, this indicates a statistically different performance.

The results indicate that, for daily and half-hourly series, the DARIMA method consistently achieves
the best forecast accuracy in terms of the mean values of MASE and MSIS, especially for long-term
forecasting. The corresponding MCB results indicate that DARIMA achieves the best-ranked performance
as well, except that it ranks second in MASE for the half-hourly frequency but without significantly
differing from the best.

In terms of hourly time series, DARIMA provides better forecasts, measured by the average MASE
and MSIS values, than ARIMA and DARIMA\_SA, but worse forecasts than ETS. However, the main comparison
we are interested in is DARIMA versus ARIMA. Moreover, the corresponding
MCB results show that DARIMA ranks third but is very close to the top two methods in
terms of MASE, while it ranks best and displays a clear gap from the remaining ones when MSIS
is used for conducting the MCB test.
A more in-depth analysis shows that DARIMA provides about $20.77\%$ more accurate interval
forecasts than ETS, based on the median of MSIS, with a smaller degree of variation.
This demonstrates that DARIMA tends to provide more stable forecasts than the competing methods.
Additionally, when there are subseries that are poorly fitted by ARIMA models, the use of DARIMA helps eliminate
the influence of outliers on the forecast performance, which can not be achieved using the
DARIMA\_SA method.

However, we observe that DARIMA may result in lower-than-nominal coverage and
yield ACD values that are higher than ARIMA, but lower than or comparable to DARIMA\_SA. The loss
of the efficiency of the estimator may be attributed to the simplified treatment of
setting $\widehat{\Sigma}_{k}$ with $\hat{\sigma}_{k}^{2} I$ in Section~\ref{sec:dlsa}.
Furthermore, DARIMA, on average, ranks higher in MSIS than ARIMA, DARIMA\_SA, and ETS as shown in
Figure~\ref{fig:sim}, thus enabling optimal decision making with a comprehensive understanding
of the uncertainty and the resulting risks. Thus, we conclude that DARIMA does not only provide stable
forecasts across different frequencies, but also achieve improved or at least comparable forecasts
compared to the benchmark methods.

\begin{table}[!ht]
  \centering
  \caption{Benchmarking the performance of DARIMA against DARIMA\_SA, ARIMA, and ETS
  with regard to MASE, MSIS and ACD. For each measure, the minimum score among the
  four methods is marked in \textbf{bold}.\protect\footnotemark}
  \resizebox{\linewidth}{!}{
  \begin{tabular}{lrrrrrrrrr}
    \toprule
            & \multicolumn{3}{c}{Daily}                                                        & \multicolumn{3}{c}{Hourly}                                                       & \multicolumn{3}{c}{Half-hourly} \\
    \cmidrule(lr){2-4} \cmidrule(lr){5-7} \cmidrule(lr){8-10}
            & \multicolumn{1}{c}{Short} & \multicolumn{1}{c}{Long} & \multicolumn{1}{c}{Total} & \multicolumn{1}{c}{Short} & \multicolumn{1}{c}{Long} & \multicolumn{1}{c}{Total} & \multicolumn{1}{c}{Short} & \multicolumn{1}{c}{Long} & \multicolumn{1}{c}{Total} \\
    \midrule
               & \multicolumn{9}{c}{MASE}                                                                                                                                                               \\
    DARIMA     & \textbf{0.700}  & \textbf{1.985}   & \textbf{1.626}  & 3.277              & 11.543              & 8.766               & \textbf{2.622}     & \textbf{8.472}      & \textbf{6.506}      \\
    DARIMA\_SA & 0.704           & 1.997            & 1.635           & $4.002^{\dagger}$  & $12.911^{\dagger}$  & $9.917^{\dagger}$   & $2.879^{\dagger}$  & $9.044^{\dagger}$   & $6.973^{\dagger}$   \\
    ARIMA      & 0.793           & 2.670            & 2.144           & 3.609              & 14.517              & 10.852              & 3.991              & 15.144              & 11.397              \\
    ETS        & 0.838           & 2.143            & 1.778           & \textbf{2.442}     & \textbf{7.764}      & \textbf{5.976}      & 8.705              & 24.663              & 19.301              \\
    \cmidrule(lr){2-10}
               & \multicolumn{9}{c}{MSIS}                                                                                                                                                               \\
    DARIMA     & \textbf{3.921}  & \textbf{11.574}  & \textbf{9.431}  & 44.671             & 200.993             & 148.469             & \textbf{24.608}    & \textbf{104.287}    & \textbf{77.515}     \\
    DARIMA\_SA & 4.027           & 11.815           & 9.634           & $74.645^{\dagger}$ & $256.492^{\dagger}$ & $195.391^{\dagger}$ & $36.263^{\dagger}$ & $134.006^{\dagger}$ & $101.164^{\dagger}$ \\
    ARIMA      & 4.683           & 20.993           & 16.426          & 62.030             & 490.435             & 346.491             & 90.331             & 710.849             & 502.355             \\
    ETS        & 5.688           & 17.713           & 14.346          & \textbf{30.671}    & \textbf{130.105}    & \textbf{96.696}     & 137.094            & 614.441             & 454.053             \\
    \cmidrule(lr){2-10}
               & \multicolumn{9}{c}{ACD}                                                                                                                                                                \\
    DARIMA     & 0.003           & 0.006            & 0.005           & 0.099              & 0.140               & 0.126               & 0.067              & 0.125               & 0.106               \\
    DARIMA\_SA & 0.007           & \textbf{0.002}   & \textbf{0.003}  & 0.131              & 0.173               & 0.159               & 0.097              & 0.150               & 0.132               \\
    ARIMA      & \textbf{0.002}  & 0.011            & 0.009           & \textbf{0.000}     & \textbf{0.001}      & \textbf{0.001}      & \textbf{0.003}     & \textbf{0.012}      & \textbf{0.009}      \\
    ETS        & 0.004           & 0.032            & 0.024           & 0.107              & 0.140               & 0.129               & 0.039              & 0.086               & 0.070               \\
    \bottomrule
  \end{tabular}}
  \label{tab:sim}
\end{table}
\footnotetext{The MASE and MSIS results of DARIMA\_SA are greatly affected by an extreme outlier
  series in the hourly and half-hourly frequencies, respectively. We present the average
  MASE and MSIS values after removing the outlier series (marked with $\dagger$) for
  hourly and half-hourly series in the table. We also provide the actual average values
  here. For the hourly series, the average MASE results across short-term, long-term, and
  all forecast horizons are $5.022\times10^{8}$, $1.036\times10^{32}$, and
  $6.877\times10^{31}$, while the MSIS values are $1.996\times10^{10}$,
  $4.118\times10^{33}$, and $2.734\times10^{33}$, respectively. For the half-hourly
  series, the average MSIS results across short-term, long-term, and all forecast horizons
  are $2.162\times10^{16}$, $1.318\times10^{59}$, and $8.750\times10^{58}$, while the MSIS
  results are $8.198\times10^{17}$, $5.007\times10^{60}$, and $3.325\times10^{60}$,
  respectively.}

\begin{figure}[!ht]
    \centering
    \includegraphics[width=\textwidth]{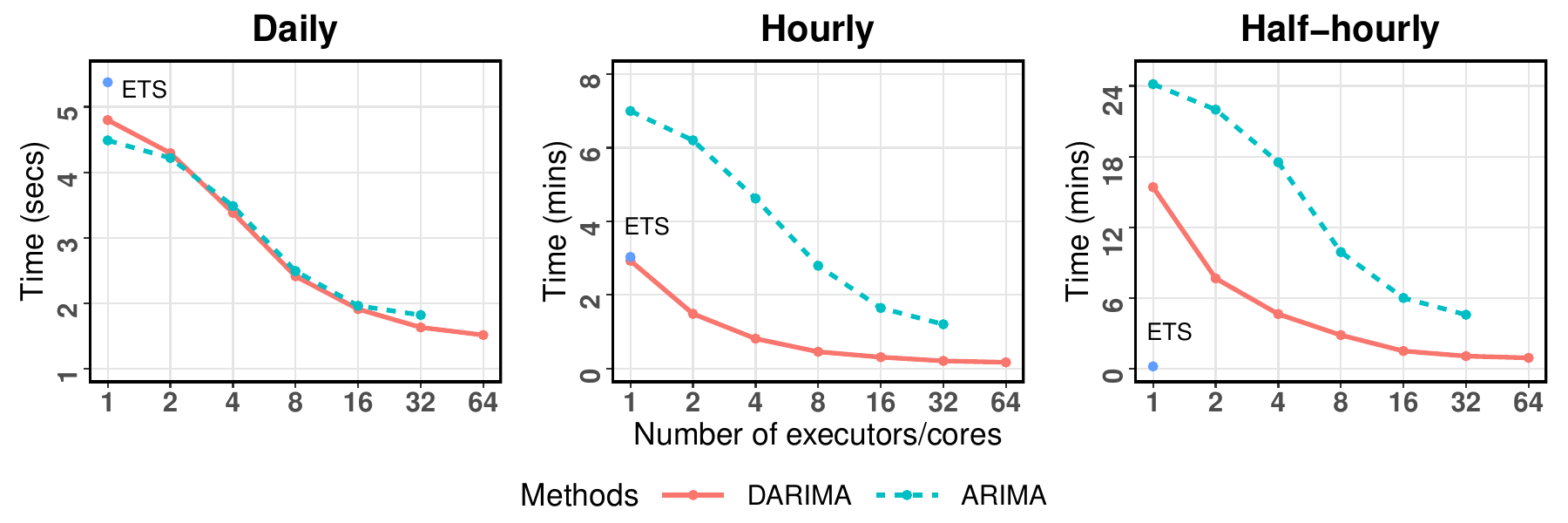}
    \caption{The average execution time of modeling DARIMA, ARIMA, and ETS for a single time series by
    using different numbers of executors/cores. There is no significant difference in the execution time
    between DARIMA and DARIMA\_SA. Note that the execution time of ETS for half-hourly series can not
    be used for comparison purposes because the \code{ets()} function in the \pkg{forecast}
    \proglang{R} package restricts seasonality to be a maximum period of $24$.
    Therefore, for half-hourly series ($m=48$), none of the ETS models contains a seasonal component and,
    thus, the ETS models produce terrible forecasts.}
    \label{fig:time_sim}
\end{figure}

In order to investigate the computational cost of our proposed method, we proceed by
visualising the computational time of DARIMA and the three benchmark methods for each data
frequency, as presented in Figure~\ref{fig:time_sim}. It should be noted that the execution
time of DARIMA\_SA is not included because there is no significant difference in the execution time
between DARIMA and DARIMA\_SA: they differ only in terms of how to combine the estimated parameters
trained on each subseries, and the time required for the combination step is negligible.
Moreover, the execution time of ETS does not change along with the number of virtual
cores used, since ETS modeling for a single time series cannot be extended to parallel processing.
We observe that, as the length of the time series data of interest increases, both DARIMA
and ARIMA take more time modeling the time series. The runtime required for both ARIMA
and DARIMA modeling decreases as the number of executors/cores increases. However,
DARIMA is computationally more efficient than ARIMA and ETS and keeps its runtime and forecasts
within a reasonable and acceptable range when using more than eight executors.

\section{Discussion}
\label{sec:discussion}

Advances in technology have given rise to increasing demand for forecasting time series
data spanning a long time interval, which is extremely challenging to achieve. Attempts to
tackle the challenge by using MapReduce technology typically focus on two mainstream
directions: combining forecasts from multiple models \citep{bendre2019time} and splitting
the multi-step forecasting problem into $H$ (forecast horizon) subproblems
\citep{galicia2018novel}. On the other hand, the statistical computation can be
implemented on a distributed system by aggregating the information about local estimators
transmitted from worker nodes \citep{fan2019distributed, zhu2021least}. The approach
results in the combined estimator proven to be statistically as efficient as the global
estimator. Inspired by the solution, this study provides a new way to forecast ultra-long
time series on a distributed system.

One of our developed framework highlights is that the distributed forecasting framework is
dedicated to averaging the DGP of subseries to develop a trustworthy global model for time
series forecasting. To this end, instead of unrealistically assuming the DGP of time
series data remains invariant over an ultra-long time period
\citep{hyndman2018forecasting}, we customize the optimization process of model parameters
for each subseries by only assuming that the DGP of subseries stays invariant over a short
period, and then aggregate these local parameters to produce the combined global model. In
this way, we provide a complete novel perspective of forecasting ultra-long time series,
with significantly improved computational efficiency.

As illustrated in Section~\ref{sec:method}, this study focuses on implementing the
distributed time series forecasting framework using general ARIMA models that allow the
inclusion of additional seasonal terms to deal with strong seasonal behavior.
Nevertheless, it is also possible to apply the framework with other statistical models,
such as state-space models, VAR models, and ETS, as a general loss function is considered.
However, special concerns should be given on how to properly convert the local estimators
to avoid the inefficiency in the combined estimators. In this work, we restrict our attention
to ARIMA models and involve a linear transformation step, in which ARIMA models trained on
subseries are converted into linear representations to avoid the stationary, causality,
and invertibility problems that may be caused by directly combining the original parameters
of ARIMA models. Similar to ARIMA models, ETS models share the virtue of allowing the trend and
seasonal components to vary over time \citep{hyndman2002state}. We hope to shed some light
on using distributed ETS models in the future.

The forecasting performance of the distributed ARIMA models is affected by various
factors. Two factors, the number of split subseries and the maximum values of model
orders, are taken into consideration as described in Section~\ref{sec:factors}. Our
results show that the number of subseries should be limited to a reasonable range to
achieve improved performance in point forecasts and prediction intervals. Specifically, we
recommend that subseries' length ranges from $500$ to $1200$ for hourly time
series. Moreover, compared to ARIMA models, a smaller maximum value of model order is
sufficient for the distributed ARIMA models to fit models for all subseries and obtain
improved forecasting results according to the combined estimators.

Many other potential factors may hold sway over the forecasting performance of our
proposed approach. For example, whether to set an overlap between successive subseries may
be a practical consideration when implementing the proposed distributed forecasting
framework. Through repeated surveys, \citet{scott1974analysis} explore the effect of
whether to overlap the random samples at each period on the estimation of population
parameters. They illustrate that considering the overlap between samples offers reductions
in the variance; they also discuss the optimum proportion of overlap. Therefore, we
believe that a study on setting overlap between successive subseries will further improve
our framework, and our framework and computer code are generally applicable to such a
scenario. To take another example, we may consider adding time-dependent weights for each
subseries when combining the local estimators delivered from a group of worker nodes. The
time-dependent weights for subseries help assign higher weights to subseries closer to the
forecast origin, while smaller weights to subseries that are further away from the
forecast origin.

\section{Conclusions}
\label{sec:conclusion}

In this paper, we propose a novel framework for ultra-long time series forecasting on a
distributed system. Unlike previous attempts in the forecasting literature, this study
facilitates distributed time series forecasting by taking a weighted average of the local
estimators delivered from worker nodes to minimize the global loss function. To this end,
an ultra-long time series spanning a long stretch of time is divided into several
subseries spanning short time periods. Specifically, in this study, we focus on
implementing our proposed framework on ARIMA models to enable the ARIMA estimation of
ultra-long time series in a distributed manner.

We investigate the performance of the distributed ARIMA models in both the real data
application and the simulations and compare the proposed approach against ARIMA models
concerning point forecasts, prediction intervals, and execution time. We find that the
distributed ARIMA models outperform ARIMA models in both point forecasting and uncertainty
estimation. The achieved performance improvements become more pronounced as the forecast
horizon increases. Finally, the comparison of execution time shows that our approach also
achieves better forecasting performance with improved computational efficiency. We also
explore various factors that may affect the forecasting performance of the distributed
ARIMA models, such as the number of split subseries, the maximum values of model orders,
overlap between successive subseries, and time-dependent weights for subseries. To further
improve the research on distributed forecasting methods, we suggest some possible research
avenues. For example, it would be meaningful to explore distributed ETS models in the
future.

\section*{Acknowledgments}

The authors are grateful to the editors and two anonymous reviewers for helpful comments
that improved the contents of the paper.

Yanfei Kang is supported the National Natural Science Foundation of China (No. 72171011)
and Feng Li is supported by the Emerging Interdisciplinary Project of
CUFE and the Beijing Universities Advanced Disciplines Initiative (No. GJJ2019163). This
research is supported by Alibaba Group through the Alibaba Innovative Research Program
and the high-performance computing (HPC) resources at Beihang University.

\printbibliography

@Article{sommer2021online,
  author    = {Sommer, Benedikt and Pinson, Pierre and Messner, Jakob W and Obst, David},
  journal   = {International Journal of Forecasting},
  title     = {Online distributed learning in wind power forecasting},
  year      = {2021},
  number    = {1},
  pages     = {205--223},
  volume    = {37},
  publisher = {Elsevier},
}

@Article{gonccalves2021critical,
  author    = {Gon{\c{c}}alves, Carla and Bessa, Ricardo J and Pinson, Pierre},
  journal   = {International Journal of Forecasting},
  title     = {A critical overview of privacy-preserving approaches for collaborative forecasting},
  year      = {2021},
  number    = {1},
  pages     = {322--342},
  volume    = {37},
  publisher = {Elsevier},
}

@Article{pan2021note,
  author  = {Rui Pan and Tunan Ren and Baishan Guo and Feng Li and Guodong and Hansheng Wang},
  journal = {Journal of Business and Economic Statistics},
  title   = {A Note on Distributed Quantile Regression by Pilot Sampling and One-Step Updating},
  year    = {2021},
  volume  = {In Press},
  doi     = {10.1080/07350015.2021.1961789},
}

@Article{zhu2021least,
  author  = {Zhu, Xuening and Li, Feng and Wang, Hansheng},
  journal = {Journal of Computational and Graphical Statistics},
  title   = {Least-Square Approximation for a Distributed System},
  year    = {2021},
  number  = {4},
  pages   = {1004-1018},
  volume  = {30},
  doi     = {10.1080/10618600.2021.1923517},
}

@Article{wang2021uncertainty,
  author    = {Wang, Xiaoqian and Kang, Yanfei and Petropoulos, Fotios and Li, Feng},
  journal   = {Journal of the Operational Research Society},
  title     = {The uncertainty estimation of feature-based forecast combinations},
  year      = {2021},
  number    = {In Press},
  doi       = {10.1080/01605682.2021.1880297},
  publisher = {Taylor & Francis},
}

@Article{petropoulos2020forecasting,
  title={Forecasting: theory and practice},
  author={Fotios Petropoulos and Daniele Apiletti and Vassilios Assimakopoulos and Mohamed Zied Babai and Devon K. Barrow and Souhaib Ben Taieb and Christoph Bergmeir and Ricardo J. Bessa and Jakub Bijak and John E. Boylan and Jethro Browell and Claudio Carnevale and Jennifer L. Castle and Pasquale Cirillo and Michael P. Clements and Clara Cordeiro and Fernando Luiz Cyrino Oliveira and Shari De Baets and Alexander Dokumentov and Joanne Ellison and Piotr Fiszeder and Philip Hans Franses and David T. Frazier and Michael Gilliland and M. Sinan Gönül and Paul Goodwin and Luigi Grossi and Yael Grushka-Cockayne and Mariangela Guidolin and Massimo Guidolin and Ulrich Gunter and Xiaojia Guo and Renato Guseo and Nigel Harvey and David F. Hendry and Ross Hollyman and Tim Januschowski and Jooyoung Jeon and Victor Richmond R. Jose and Yanfei Kang and Anne B. Koehler and Stephan Kolassa and Nikolaos Kourentzes and Sonia Leva and Feng Li and Konstantia Litsiou and Spyros Makridakis and Gael M. Martin and Andrew B. Martinez and Sheik Meeran and Theodore Modis and Konstantinos Nikolopoulos and Dilek Önkal and Alessia Paccagnini and Anastasios Panagiotelis and Ioannis Panapakidis and Jose M. Pavía and Manuela Pedio and Diego J. Pedregal and Pierre Pinson and Patrícia Ramos and David E. Rapach and J. James Reade and Bahman Rostami-Tabar and Michał Rubaszek and Georgios Sermpinis and Han Lin Shang and Evangelos Spiliotis and Aris A. Syntetos and Priyanga Dilini Talagala and Thiyanga S. Talagala and Len Tashman and Dimitrios Thomakos and Thordis Thorarinsdottir and Ezio Todini and Juan Ramón Trapero Arenas and Xiaoqian Wang and Robert L. Winkler and Alisa Yusupova and Florian Ziel},
  year={2022},
  journal={International Journal of Forecasting}
}

@Article{LI2020imaging,
  author  = {Li, Xixi and Kang, Yanfei and Li, Feng},
  journal = {Expert Systems with Applications},
  title   = {Forecasting with time series imaging},
  year    = {2020},
  pages   = {113680},
  volume  = {160},
}

@Article{montero2020fforma,
  author    = {Montero-Manso, Pablo and Athanasopoulos, George and Hyndman, Rob J and Talagala, Thiyanga S},
  journal   = {International Journal of Forecasting},
  title     = {{FFORMA}: {Feature-based} forecast model averaging},
  year      = {2020},
  number    = {1},
  pages     = {86--92},
  volume    = {36},
  publisher = {Elsevier},
}

@Article{makridakis2020m4,
  author    = {Makridakis, Spyros and Spiliotis, Evangelos and Assimakopoulos, Vassilios},
  journal   = {International Journal of Forecasting},
  title     = {{The M4 Competition}: 100,000 time series and 61 forecasting methods},
  year      = {2020},
  number    = {1},
  pages     = {54--74},
  volume    = {36},
  publisher = {Elsevier},
}

@Article{anil2020apache,
  author  = {Anil, Robin and Capan, Gokhan and Drost-Fromm, Isabel and Dunning, Ted and Friedman, Ellen and Grant, Trevor and Quinn, Shannon and Ranjan, Paritosh and Schelter, Sebastian and Y{\i}lmazel, {\"O}zg{\"u}r},
  journal = {Journal of Machine Learning Research},
  title   = {{Apache Mahout}: Machine Learning on Distributed Dataflow Systems},
  year    = {2020},
  number  = {127},
  pages   = {1--6},
  volume  = {21},
}

@Article{das2020predictive,
  author    = {Das, Srinjoy and Politis, Dimitris N},
  journal   = {Journal of the American Statistical Association},
  title     = {Predictive inference for locally stationary time series with an application to climate data},
  year      = {2020},
  volume = 116,
  number = 534,
  pages     = {919-934},
  publisher = {Taylor \& Francis},
}

@Article{kang2020gratis,
  author  = {Kang, Yanfei and Hyndman, Rob J and Li, Feng},
  journal = {Statistical Analysis and Data Mining},
  title   = {{GRATIS}: {GeneRAting TIme Series} with diverse and controllable characteristics},
  year    = {2020},
  pages   = {354-376},
  volume  = {13},
}

@Misc{spark,
  author = {{Apache Software Foundation}},
  title  = {Apache Spark},
  year   = {2020},
  url    = {https://spark.apache.org},
}

@Article{volgushev2019distributed,
  author    = {Volgushev, Stanislav and Chao, Shih-Kang and Cheng, Guang},
  journal   = {Annals of Statistics},
  title     = {Distributed inference for quantile regression processes},
  year      = {2019},
  number    = {3},
  pages     = {1634--1662},
  volume    = {47},
  publisher = {Institute of Mathematical Statistics},
}

@Article{fan2019distributed,
  author    = {Fan, Jianqing and Wang, Dong and Wang, Kaizheng and Zhu, Ziwei},
  journal   = {Annals of Statistics},
  title     = {Distributed estimation of principal eigenspaces},
  year      = {2019},
  pages     = {3009-3031},
  volume    = {47},
  publisher = {NIH Public Access},
}

@Article{chen2019quantile,
  author    = {Chen, Xi and Liu, Weidong and Zhang, Yichen},
  journal   = {Annals of Statistics},
  title     = {Quantile regression under memory constraint},
  year      = {2019},
  number    = {6},
  pages     = {3244--3273},
  volume    = {47},
  publisher = {Institute of Mathematical Statistics},
}

@Article{bendre2019time,
  author    = {Bendre, Mininath and Manthalkar, Ramchandra},
  journal   = {Expert Systems with Applications},
  title     = {Time series decomposition and predictive analytics using MapReduce framework},
  year      = {2019},
  pages     = {108--120},
  volume    = {116},
  publisher = {Elsevier},
}

@Article{hong2019global,
  author    = {Hong, Tao and Xie, Jingrui and Black, Jonathan},
  journal   = {International Journal of Forecasting},
  title     = {{Global} energy forecasting competition 2017: {Hierarchical} probabilistic load forecasting},
  year      = {2019},
  number    = {4},
  pages     = {1389--1399},
  volume    = {35},
  publisher = {Elsevier},
}

@Article{Jordan2019,
  author    = {Jordan, Michael I and Lee, Jason D and Yang, Yun},
  journal   = {Journal of the American Statistical Association},
  title     = {Communication-Efficient Distributed Statistical Inference},
  year      = {2019},
  number    = {526},
  pages     = {668-681},
  volume    = {114},
  publisher = {Taylor & Francis},
}

@Article{galicia2018novel,
  author    = {Galicia, Antonio and Torres, Jos\'{e} F and Mart\'{i}nez-\'{A}lvarez, Francisco and Troncoso, A},
  journal   = {Information Sciences},
  title     = {A novel spark-based multi-step forecasting algorithm for big data time series},
  year      = {2018},
  pages     = {800--818},
  volume    = {467},
  publisher = {Elsevier},
}

@Article{talavera2018big,
  author    = {Talavera-Llames, R and P\'{e}rez-Chac\'{o}n, Rub\'{e}n and Troncoso, A and Mart\'{i}nez-\'{A}lvarez, Francisco},
  journal   = {Knowledge-Based Systems},
  title     = {Big data time series forecasting based on nearest neighbours distributed computing with {Spark}},
  year      = {2018},
  pages     = {12--25},
  volume    = {161},
  publisher = {Elsevier},
}

@Book{hyndman2018forecasting,
  author    = {Hyndman, Rob J and Athanasopoulos, George},
  publisher = {OTexts},
  title     = {Forecasting: principles and practice},
  year      = {2021},
  edition = "3rd",
  url = {https://OTexts.com/fpp3}
}

@Article{shang2017grouped,
  author    = {Shang, Han Lin and Hyndman, Rob J},
  journal   = {Journal of Computational and Graphical Statistics},
  title     = {Grouped functional time series forecasting: {An} application to age-specific mortality rates},
  year      = {2017},
  number    = {2},
  pages     = {330--343},
  volume    = {26},
  publisher = {Taylor \& Francis},
}

@Article{lee2015communication,
  author  = {Lee, Jason D and Liu, Qiang and Sun, Yuekai and Taylor, Jonathan E},
  journal = {Journal of Machine Learning Research},
  title   = {Communication-efficient sparse regression},
  year    = {2017},
  number  = {5},
  pages   = {1-30},
  volume  = {18},
}

@InProceedings{wang2017efficient,
  author       = {Wang, Jialei and Kolar, Mladen and Srebro, Nathan and Zhang, Tong},
  booktitle    = {International Conference on Machine Learning},
  title        = {Efficient distributed learning with sparsity},
  year         = {2017},
  organization = {PMLR},
  pages        = {3636--3645},
}

@Article{coluccia2016bayesian,
  author    = {Coluccia, Angelo and Notarstefano, Giuseppe},
  journal   = {IEEE Transactions on Signal Processing},
  title     = {A Bayesian framework for distributed estimation of arrival rates in asynchronous networks},
  year      = {2016},
  number    = {15},
  pages     = {3984--3996},
  volume    = {64},
  publisher = {IEEE},
}

@Article{meng2016mllib,
  author    = {Meng, Xiangrui and Bradley, Joseph and Yavuz, Burak and Sparks, Evan and Venkataraman, Shivaram and Liu, Davies and Freeman, Jeremy and Tsai, DB and Amde, Manish and Owen, Sean and others},
  journal   = {Journal of Machine Learning Research},
  title     = {{MLlib}: {Machine learning in apache spark}},
  year      = {2016},
  number    = {1},
  pages     = {1235--1241},
  volume    = {17},
  publisher = {JMLR. org},
}

@Book{brockwell2016introduction,
  author    = {Brockwell, Peter J and Davis, Richard A},
  publisher = {Switzerland: Springer International Publishing},
  title     = {Introduction to time series and forecasting},
  year      = {2016},
}

@Article{zhang2015divide,
  author    = {Zhang, Yuchen and Duchi, John and Wainwright, Martin},
  journal   = {Journal of Machine Learning Research},
  title     = {Divide and conquer kernel ridge regression: A distributed algorithm with minimax optimal rates},
  year      = {2015},
  number    = {1},
  pages     = {3299--3340},
  volume    = {16},
  publisher = {JMLR. org},
}

@InProceedings{maclaurin2015firefly,
  author    = {Maclaurin, Dougal and Adams, Ryan Prescott},
  booktitle = {Twenty-Fourth International Joint Conference on Artificial Intelligence},
  title     = {{Firefly Monte Carlo}: {Exact MCMC} with subsets of data},
  year      = {2015},
}

@Book{box2015time,
  author    = {Box, George EP and Jenkins, Gwilym M and Reinsel, Gregory C and Ljung, Greta M},
  publisher = {John Wiley \& Sons},
  title     = {Time series analysis: forecasting and control},
  year      = {2015},
}

@Article{liu2014distributed,
  author  = {Liu, Qiang and Ihler, Alexander},
  journal = {Advances in Neural Information Processing Systems},
  title   = {Distributed estimation, information loss and exponential families},
  year    = {2014},
  pages   = {1098–1106},
}

@Article{calheiros2014workload,
  author    = {Calheiros, Rodrigo N and Masoumi, Enayat and Ranjan, Rajiv and Buyya, Rajkumar},
  journal   = {IEEE Transactions on Cloud Computing},
  title     = {Workload prediction using {ARIMA} model and its impact on cloud applications’ {QoS}},
  year      = {2014},
  number    = {4},
  pages     = {449--458},
  volume    = {3},
  publisher = {IEEE},
}

@Article{kleiner2014scalable,
  author    = {Kleiner, Ariel and Talwalkar, Ameet and Sarkar, Purnamrita and Jordan, Michael I},
  journal   = {Journal of the Royal Statistical Society: Series B (Statistical Methodology)},
  title     = {A scalable bootstrap for massive data},
  year      = {2014},
  number    = {4},
  pages     = {795--816},
  volume    = {76},
  publisher = {Wiley Online Library},
}

@InProceedings{shamir2014communication,
  author       = {Shamir, Ohad and Srebro, Nati and Zhang, Tong},
  booktitle    = {International Conference on Machine Learning},
  title        = {Communication-efficient distributed optimization using an approximate {Newton-type} method},
  year         = {2014},
  organization = {PMLR},
  pages        = {1000--1008},
}

@InProceedings{li2014rolling,
  author       = {Li, Lei and Noorian, Farzad and Moss, Duncan JM and Leong, Philip HW},
  booktitle    = {{Proceedings of the 2014 IEEE 15th International Conference on Information Reuse and Integration (IEEE IRI 2014)}},
  title        = {Rolling window time series prediction using {MapReduce}},
  year         = {2014},
  organization = {IEEE},
  pages        = {757--764},
}

@Article{wang2013parallelizing,
  author  = {Wang, Xiangyu and Dunson, David B},
  journal = {arXiv preprint arXiv:1312.4605},
  title   = {Parallelizing MCMC via Weierstrass sampler},
  year    = {2013},
}

@Article{mirko2013hadoop,
  author    = {K\"{a}mpf, Mirko and Kantelhardt, Jan W},
  journal   = {International Journal of Computer Applications},
  title     = {{Hadoop.TS}: large-scale time-series processing},
  year      = {2013},
  number    = {17},
  pages     = {1-8},
  volume    = {74},
  publisher = {Foundation of Computer Science},
}

@Article{zhang2013communication,
  author    = {Zhang, Yuchen and Duchi, John C and Wainwright, Martin J},
  journal   = {Journal of Machine Learning Research},
  title     = {Communication-efficient algorithms for statistical optimization},
  year      = {2013},
  number    = {1},
  pages     = {3321--3363},
  volume    = {14},
  publisher = {JMLR. org},
}

@Article{fan2011high,
  author    = {Fan, Jianqing and Liao, Yuan and Mincheva, Martina},
  journal   = {Annals of Statistics},
  title     = {High dimensional covariance matrix estimation in approximate factor models},
  year      = {2011},
  number    = {6},
  pages     = {3320},
  volume    = {39},
  publisher = {NIH Public Access},
}

@Article{hyndman2011optimal,
  author    = {Hyndman, Rob J and Ahmed, Roman A and Athanasopoulos, George and Shang, Han Lin},
  journal   = {Computational Statistics \& Data Analysis},
  title     = {Optimal combination forecasts for hierarchical time series},
  year      = {2011},
  number    = {9},
  pages     = {2579--2589},
  volume    = {55},
  publisher = {Elsevier},
}

@article{boyd2011distributed,
  author    = {Boyd, Stephen and Parikh, Neal and Chu, Eric},
  journal = {Foundations and Trends in Machine Learning},
  volume = 3,
  number = 1,
  pages = {1-122},
  title     = {Distributed optimization and statistical learning via the alternating direction method of multipliers},
  year      = {2011},
}

@Article{suchard2010understanding,
  author    = {Suchard, Marc A and Wang, Quanli and Chan, Cliburn and Frelinger, Jacob and Cron, Andrew and West, Mike},
  journal   = {Journal of Computational and Graphical Statistics},
  title     = {Understanding {GPU} programming for statistical computation: {Studies} in massively parallel massive mixtures},
  year      = {2010},
  number    = {2},
  pages     = {419--438},
  volume    = {19},
  publisher = {Taylor \& Francis},
}

@Book{yuen2010bayesian,
  author    = {Yuen, K. V.},
  publisher = {John Wiley \& Sons},
  address = {Singapore},
  title     = {Bayesian methods for structural dynamics and civil engineering},
  year      = {2010},
}

@Article{fan2008high,
  author    = {Fan, Jianqing and Fan, Yingying and Lv, Jinchi},
  journal   = {Journal of Econometrics},
  title     = {High dimensional covariance matrix estimation using a factor model},
  year      = {2008},
  number    = {1},
  pages     = {186--197},
  volume    = {147},
  publisher = {Elsevier},
}

@Article{fan2008statistical,
  author    = {Fan, Jianqing and Zhang, Wenyang},
  journal   = {Statistics and its Interface},
  title     = {Statistical methods with varying coefficient models},
  year      = {2008},
  pages     = {179-195},
  volume    = {1},
  publisher = {NIH Public Access},
}

@Article{gneiting2007strictly,
  author    = {Gneiting, Tilmann and Raftery, Adrian E},
  journal   = {Journal of the American Statistical Association},
  title     = {Strictly proper scoring rules, prediction, and estimation},
  year      = {2007},
  number    = {477},
  pages     = {359--378},
  volume    = {102},
  publisher = {Taylor \& Francis},
}

@Book{tanenbaum2007distributed,
  author    = {Tanenbaum, Andrew S and Van Steen, Maarten},
  publisher = {Prentice-Hall},
  title     = {Distributed systems: principles and paradigms},
  year      = {2007},
}

@Article{hyndman2006another,
  author    = {Hyndman, Rob J and Koehler, Anne B},
  journal   = {International Journal of Forecasting},
  title     = {Another look at measures of forecast accuracy},
  year      = {2006},
  number    = {4},
  pages     = {679--688},
  volume    = {22},
  publisher = {Elsevier},
}

@Article{koning2005m3,
  author    = {Koning, Alex J and Franses, Philip Hans and Hibon, Michele and Stekler, Herman O},
  journal   = {International Journal of Forecasting},
  title     = {The {M3} competition: {Statistical tests of the results}},
  year      = {2005},
  number    = {3},
  pages     = {397--409},
  volume    = {21},
  publisher = {Elsevier},
}

@Book{tsay2005analysis,
  author    = {Tsay, Ruey S},
  publisher = {John Wiley \& Sons},
  address = {Hoboken, NJ, USA},
  title     = {Analysis of financial time series},
  year      = {2010},
  edition = "3rd"
}

@InProceedings{ghemawat2003google,
  author    = {Ghemawat, Sanjay and Gobioff, Howard and Leung, Shun-Tak},
  booktitle = {{Proceedings of the nineteenth ACM symposium on Operating systems principles}},
  title     = {The {Google} file system},
  year      = {2003},
  pages     = {29--43},
}

@Article{hyndman2002state,
  author    = {Hyndman, Rob J and Koehler, Anne B and Snyder, Ralph D and Grose, Simone},
  journal   = {International Journal of Forecasting},
  title     = {A state space framework for automatic forecasting using exponential smoothing methods},
  year      = {2002},
  number    = {3},
  pages     = {439--454},
  volume    = {18},
  publisher = {Elsevier},
}

@Article{canova1995seasonal,
  author    = {Canova, Fabio and Hansen, Bruce E},
  journal   = {Journal of Business \& Economic Statistics},
  title     = {{Are} seasonal patterns constant over time? {A} test for seasonal stability},
  year      = {1995},
  number    = {3},
  pages     = {237--252},
  volume    = {13},
  publisher = {Taylor \& Francis Group},
}

@Article{makridakis1993accuracy,
  author    = {Makridakis, Spyros},
  journal   = {International Journal of Forecasting},
  title     = {Accuracy measures: theoretical and practical concerns},
  year      = {1993},
  number    = {4},
  pages     = {527--529},
  volume    = {9},
  publisher = {Elsevier},
}

@Article{kwiatkowski1992testing,
  author  = {Kwiatkowski, Denis and Phillips, Peter CB and Schmidt, Peter and Shin, Yongcheol and others},
  journal = {Journal of Econometrics},
  title   = {Testing the null hypothesis of stationarity against the alternative of a unit root},
  year    = {1992},
  number  = {1-3},
  pages   = {159--178},
  volume  = {54},
}

@Article{scott1974analysis,
  author    = {Scott, AJ and Smith, TMF},
  journal   = {Journal of the American Statistical Association},
  title     = {Analysis of repeated surveys using time series methods},
  year      = {1974},
  number    = {347},
  pages     = {674--678},
  volume    = {69},
  publisher = {Taylor \& Francis Group},
}

@ARTICLE{Hyndman2008b,
  author    = {Hyndman, Rob J and Khandakar, Yeasmin},
  journal   = {Journal of Statistical Software},
  title     = {Automatic Time Series Forecasting: The forecast Package for {R}},
  publisher = {jstatsoft.org},
  volume    = {27},
  pages     = {1--22},
  year      = {2008},
  doi       = {10.18637/jss.v027.i03}
}
\clearpage
\setcounter{table}{0}
\setcounter{figure}{0}

\appendix
\pagenumbering{arabic}

\section{Background of Distributed Systems}
\label{sec:distr}

A distributed system, usually used for distributed computing, is a system with a group of
interacting computing nodes connected by a network \citep{tanenbaum2007distributed}. These
autonomous computers share resources, work together, and coordinate their activities to
fulfill specified tasks, just like a single computer via a MapReduce framework. When
dealing with large-scale problems, distributed systems provide a new solution that sends
the computing code to each computer node where data are also distributed stored. The
MapReduce is short for the ``move-code-to-data'' computing architecture that enables us to
scale horizontally by adding more computing nodes, rather than scale vertically, by
upgrading a single node's hardware.

Inspired by the Google File System paper \citep{ghemawat2003google} that describes
Google's algorithm for distributed data-intensive applications,
Hadoop ecosystem has been developed in the data science community
as an open-source platform that allows for the
distributed storage and processing of large-scale data sets. Such an ecosystem is the
\emph{de-facto} standard for large scale distributed computing in data analytics
sectors. Nonetheless, the existing distributed systems equip with machine learning libraries \citep{meng2016mllib,anil2020apache} but
lack forecasting support. Forecasters have to make unrealistic
independence assumptions for modeling large scale time series data to fit in the
ecosystem. Developing and integrating forecasting methods into such distributed systems has
great potential.

Apache Spark~\citep{spark} is the most popular distributed execution engine used for big
data processing in the distributed ecosystem. With in-memory processing, Spark does not
spend excess time moving data in and out of the disk, which achieves significantly faster
(up to $100$ times) computation. Besides, Spark supports computer languages (\eg , Java, Scala, R, and
Python) that are widely used in the machine learning and forecasting domains, making it
developer-friendly. Spark also offers a stack of libraries, such as \pkg{MLlib} for machine
learning, Spark Streaming for real-time processing, Spark SQL for interactive queries, and
GraphX for graph processing, which provides easy-to-use analytics in many cases.

\section{Pseudocode for Distributed Forecasting}
\label{sec:dfalg}

This section provides the pseudocode for Mapper and Reducer of the proposed distributed
forecasting approach in Section~\ref{sec:method}.

\begin{algorithm}[!htb]
  \renewcommand{\algorithmicrequire}{\textbf{Input:}}
  \renewcommand{\algorithmicensure}{\textbf{Output:}}
  \caption{Map function for the distributed time series forecasting. }
  \label{alg:map}
  \begin{algorithmic}
    \Require ~ $\big\langle key, value \big\rangle$
    \Ensure ~ $\big\langle key, ( \widehat{\theta}, \widehat{\Sigma} ) \big\rangle$ \\
  \end{algorithmic}
  \begin{algorithmic}[1]
    \State \textbf{Start}
    \State $y_t \leftarrow \text{time series data}$
    \State $T \leftarrow \text{time series length}$
    \State $K \leftarrow \text{number of split subseries}$
    \State $\text{index} \leftarrow \text{index vector of the assigned subseries}$
    \State
    \State $n = \text{floor}(T/K)$
    \For{$i$ in index}
       \State $\text{lbound} = n \times (i-1) + 1$
       \State $\text{ubound} = \text{ifelse} (i \geq K, T, n \times i)$
       \State $y = y_t [\text{lbound}, \text{ubound}]$ \Comment{Step $1$}
       \State $\text{fit} = \text{model} ( y ) $ \Comment{Step $2$}
       \State $\text{fit}^{'} = \text{model.to.linear} ( \text{fit} ) $ \Comment{Step $3$}
       \State $\widehat{\theta} = \text{fit}^{'}.\text{coef}$
       \State $\widehat{\Sigma} = \text{fit}^{'}.\text{var.coef}$
    \EndFor
    \State \textbf{Stop}
  \end{algorithmic}
\end{algorithm}

\begin{algorithm}[!htb]
  \renewcommand{\algorithmicrequire}{\textbf{Input:}}
  \renewcommand{\algorithmicensure}{\textbf{Output:}}
  \caption{Reduce function for the distributed time series forecasting. }
  \label{alg:reduce}
  \begin{algorithmic}
    \Require ~ $\big\langle \text{key}, \text{list} ( \widehat{\theta}, \widehat{\Sigma} ) \big\rangle$
    \Ensure ~ $\big\langle \text{key}, \text{forecvalues} \big\rangle$ \\
  \end{algorithmic}
  \begin{algorithmic}[1]
    \State \textbf{Start}
    \State $y_t \leftarrow \text{time series data}$
    \State $H \leftarrow \text{forecast horizon}$
    \State $level \leftarrow \text{confidence levels for prediction intervals}$
    \State
    \State $\widetilde{\text{fit}} = \text{Comb.method} \big( \text{list} ( \widehat{\theta}, \widehat{\Sigma} ) \big)$  \Comment{Step $4$}
    \State $\text{forec} = \text{forecast} \big( \widetilde{\text{fit}}, y_t, H, \text{level} \big)$ \Comment{Step $5$}
    \State $\text{pred} = \text{forec.mean}$ \Comment{point forecasts}
    \State $\text{lower} = \text{forec.lower}$ \Comment{lower bound of prediction intervals}
    \State $\text{upper} = \text{forec.upper}$ \Comment{upper bound of prediction intervals}
    \State \textbf{Stop}
  \end{algorithmic}
\end{algorithm}

\end{document}